\documentclass[twocolumn]{aastex62}
\pdfoutput=1
\usepackage{amsmath,amstext}
\usepackage[T1]{fontenc}
\usepackage[figure,figure*]{hypcap}
\usepackage{amsmath,amssymb}
\usepackage{lineno}
\usepackage{hyperref}
\usepackage[nameinlink]{cleveref}
\makeatletter
\usepackage{etoolbox}
\patchcmd\H@refstepcounter{\protected@edef}{\protected@xdef}{}{}
\makeatother

\usepackage{appendix}

\graphicspath{{./}{figures_new/}}
\received{\today}
\revised{\today}
\accepted{\today}
\submitjournal{ApJ}
\newlength{\colwidth}

\newcommand{\hinv}{\ensuremath{h^{-1}}}
\newcommand{\mpc}{\ensuremath{\rm{Mpc}}}
\newcommand{\mpch}{\ensuremath{h^{-1}\rm{Mpc}}}

\newcommand{\mstar}{\ensuremath{M^*}}
\newcommand{\vmpeak}{\ensuremath{v_{\rm Mpeak}}}
\newcommand{\mpeak}{\ensuremath{M_{\rm peak}}}

\newcommand{\rockstar}{\texttt{Rockstar}}

\newcommand{\nbody}{$N$-body}
\newcommand{\msol}{\ensuremath{{\rm M}_{\solar}}}
\newcommand{\multipoles}{\ensuremath{\hat{\xi}_{0/2}}}

\newcommand{\solar}{\ensuremath{_{\mathord\odot}}}
\newcommand{\hmpc}{\ifmmode{h^{-1}{\rm Mpc}}\;\else${h^{-1}}${\rm Mpc}\fi}
\newcommand{\hMpc}{\ifmmode{h^{-1}{\rm Mpc}}\;\else${h^{-1}}${\rm Mpc}\fi}
\newcommand{\hGpc}{\ifmmode{h^{-1}{\rm Gpc}}\;\else${h^{-1}}${\rm Gpc}\fi}
\newcommand{\hkpc}{\ifmmode{h^{-1}{\rm kpc}}\;\else${h^{-1}}${\rm kpc}\fi}

\newcommand{\dvmax}{\ensuremath{\Delta v_{\rm max}}}
\newcommand{\rhalo}{\ensuremath{R_{\rm h}}}

\newcommand{\mr}{\ifmmode{M_r}\;\else$M_r$\fi}
\newcommand{\mcut}{\ensuremath{M_{\rm cut}}}
\newcommand{\healpix}{\texttt{HealPix}}

\AtBeginEnvironment{appendices}{\crefalias{section}{appendix}}

\shorttitle{RSD with AM}
\shortauthors{DeRose et al.}

\begin{document}

\title{Modeling Redshift-Space Clustering with Abundance Matching}
\author[0000-0002-0728-0960]{Joseph DeRose}\thanks{jderose@lbl.gov}
\affiliation{Lawrence Berkeley National Laboratory, 1 Cyclotron Road, Berkeley, CA 93720, USA}
\author[0000-0001-7774-2246]{Matthew R. Becker}
\affiliation{High-Energy Physics Division, Argonne National Laboratory, Lemont, IL 60439, USA}
\author[0000-0003-2229-011X]{Risa H. Wechsler}
\affiliation{Department of Physics, Stanford University, 382 Via Pueblo Mall, Stanford, CA 94305, USA}
\affiliation{Kavli Institute for Particle Astrophysics \& Cosmology, P. O. Box 2450, Stanford University, Stanford, CA 94305, USA}
\affiliation{SLAC National Accelerator Laboratory, Menlo Park, CA 94025, USA}

\begin{abstract}
We explore the degrees of freedom required to jointly fit projected and redshift-space clustering of galaxies selected in three bins of stellar mass from the Sloan Digital Sky Survey Main Galaxy Sample (SDSS MGS) using a subhalo abundance matching (SHAM) model. We employ emulators for relevant clustering statistics in order to facilitate our analysis, leading to large speed gains with minimal loss of accuracy. We are able to simultaneously fit the projected and redshift-space clustering of the two most massive galaxy samples that we consider with just two free parameters: scatter in stellar mass at fixed SHAM proxy and the dependence of the SHAM proxy on dark matter halo concentration. We find some evidence for models that include velocity bias, but including orphan galaxies improves our fits to the lower mass samples significantly. We also model the clustering signals of specific star formation rate (SSFR) selected samples using conditional abundance matching (CAM). We obtain acceptable fits to projected and redshift-space clustering as a function of SSFR and stellar mass using two CAM variants, although the fits are worse than for stellar mass selected samples alone. By incorporating non-unity correlations between the CAM proxy and SSFR we are able to resolve previously identified discrepancies between CAM predictions and SDSS observations of the environmental dependence of quenching for isolated central galaxies.

\end{abstract}


\section{Introduction} \label{sec:intro}
Our theoretical understanding of galaxy formation has advanced significantly with the advent of high resolution $N$-body simulations that are capable of resolving substructure within dark matter halos. Shortly after the the first such simulations were possible, so called subhalo abundance matching (SHAM) models were devised to take advantage of this newfound ability \citep{Kravtsov2004, ValeOstriker2004, Conroy2006}. These models place simulated galaxies directly on resolved substructure, and posit that the stellar mass or luminosity of a galaxy is approximately monotonically related to the mass or velocity of the dark matter (sub)halo hosting that galaxy. Despite their simplicity, SHAM models are able to reproduce a broad range of galaxy spatial statistics, including projected two-point clustering, conditional luminosity functions, and radial profiles of galaxies within halos \citep{Reddick2013, Hearin2013b, Saito2015, Lehmann2017}. In this work, we demonstrate that SHAM models can also fit redshift-space clustering measurements.

Empirical models such as SHAM now form the basis of many wide-field galaxy survey simulations because of their ability to predict clustering over a broad range of scales and redshifts for a variety of galaxy samples \citep{DeRose2018, Korytov2019}. The SHAM approach also holds promise as a forward model for cosmological analyses that employ galaxy clustering statistics. Other simulation-based models such as halo occupation distribution (HOD) models \citep{Seljak2002, BerlindWeinberg, Bullock2003} must assume analytic models for the number of galaxies in a halo, $N$, as a function that halo's mass, $P(N|M)$, and a phase-space distribution of galaxies within halos, $P(\mathbf{x}, \mathbf{v}|M,c)$, that again depends on the host halo's mass, $M$, as well as its  concentration, $c$. Such models are able to fit galaxy projected and redshift-space clustering into the one-halo regime \citep{Reid2014,Zhai2018,Lange2021}, but usually use five to ten parameters in doing so, making SHAM a potentially more efficient parameterization.

While SHAM has been touted as a highly predictive model for galaxy clustering, questions remain about its ability to fit observables at the statistical precision afforded by modern galaxy surveys. The highest signal-to-noise galaxy clustering measurements currently available come from galaxy spectroscopic redshift surveys; to make predictions for the observables from such surveys, models must account for both galaxy positions and line-of-sight velocities. A number of investigations have been conducted into the ability of SHAM to fit redshift-space galaxy clustering measurements from these surveys. \citet{Saito2015} showed that SHAM combined with a model for stellar mass incompleteness as a function of galaxy color could fit projected clustering measurements from the BOSS CMASS sample. \citet{Yamamoto2015} compared the predictions of a SHAM model using subhalo maximum circular velocity as a proxy for galaxy absolute magnitude to redshift-space clustering statistics measured from the Sloan Digital Sky Survey Main Galaxy Sample (SDSS MGS), and found reasonable agreement, but no quantitative goodness-of-fit statistics were provided. \citet{Guo2016} performed a more quantitative comparison, investigating the ability of SHAM and HOD models to fit redshift-space distortion (RSD) measurements.  They concluded that SHAM models with a single free parameter governing scatter between the SHAM proxy and galaxy absolute magnitude were unable to simultaneously fit projected and redshift-space clustering. Recently \citet{Contreras2020} found that a SHAM model with parameters governing subhalo artificial disruption and galaxy assembly bias in addition to a parameter governing the scatter between SHAM proxy and stellar mass were able to fit RSD measurements in Illustris TNG \citep{Springel2018, Nelson2018, Marinacci2018, Naiman2018, Pillipich2018}, and projected clustering statistics from SDSS. In this work, we perform an extensive study of SHAM's ability to fit RSD measurements from the SDSS MGS, applying quantitative goodness-of-fit metrics and investigating how SHAM extensions such as orphan galaxies and velocity bias affect these fits. The effect of orphan galaxies in SHAM was also explored in \citet{Contreras2021}.

A major drawback of commonly used SHAM approaches are their inability to model galaxy samples that are incomplete in galaxy luminosity or stellar mass. Nearly all samples used for modern cosmology analyses have this property. For example, the main sample used for cosmological constraints in the Baryon Acoustic Oscillation Survey (BOSS) is the CMASS sample \citep{Dawson2013}. CMASS, while approximately complete in stellar mass, has incompleteness that is a function of galaxy star formation rate (SFR) \citep{Leauthaud2016,Saito2015}. Galaxy clustering is highly dependent on SFR \citep{Zehavi2011}, so SFR-dependent incompleteness imparts a bias in the CMASS clustering with respect to the clustering of a stellar-mass-complete sample that simple implementations of SHAM cannot model. Upcoming surveys will pose similar or even more severe sample incompleteness problems as many upcoming spectroscopic surveys are designed to target galaxies with strong emission lines \citep{Aghamousa2016, Laureijs2011, Dore2019}. Thus, methods for incorporating such selections in the SHAM framework must be developed if the approach is to be used to model these samples.

Significant progress has been made in modeling the dependence of galaxy clustering on SFR. \citet{Hearin2013} introduced an extension to SHAM, dubbed age matching, that correlates galaxy $z=0.1$-frame $g-r$ color at fixed $r$-band luminosity with dark matter subhalo formation time. They showed that age matching reproduces the dependence of projected clustering on galaxy luminosity and color at high accuracy with no free parameters other than the choice of proxies used for the SHAM and age-matching procedures. A similar method was also put forth by \citet{Masaki2013} with comparable results. Generalizations of these first models, now referred to as conditional abundance matching (CAM) models \citep{Hearin2014, Watson2015}, have been tested against SDSS galaxy-galaxy lensing as well as group and galaxy cluster statistics with encouraging results. 

CAM models have also been applied to RSD measurements in the past with little success: \citet{Yamamoto2015} constrained the models presented in \citet{Masaki2013} against SDSS MGS redshift-space clustering finding that none of their CAM models fit the data well when considering chi-squared tests. In the second part of this work we confront CAM with RSD measurements for samples split by stellar mass and SSFR. We focus on two new CAM models, one based on subhalo's accretion rate, and another that uses distance to the nearest host halo above a set mass threshold, examining the goodness-of-fit of these models when used in conjunction with a model for orphan galaxies.

The structure of this paper is as follows. In \cref{sec:obs} we describe the SDSS galaxy samples used in this work and the methodology used to measure RSD statistics including how we account for fiber collisions. In \cref{sec:sims} we describe the simulation employed in this work, which was run at the best fit cosmology from \citet{Planck2015}. \Cref{sec:models} introduces the details of our SHAM and CAM models, including how these models are implemented in our simulation. \Cref{sec:base_sham} presents the results of our SHAM fits to SDSS MGS RSD data and how these fits are impacted by the inclusion of orphan galaxies and velocity bias. In \cref{sec:cam} we fit two CAM models to SDSS MGS RSD measurements as a function of stellar mass and specific star formation rate (SSFR), and use posterior predictive distributions from these fits in order to perform a detailed comparison of these models. In \cref{sec:conclusions} we summarize our main results and conclude by discussing future directions of investigation.

\section{Observational Data and Clustering Measurements}
\label{sec:obs}

In this work we make use of the NYU Value Added Galaxy Catalog (VAGC) \citep{Blanton2005}, which is constructed from SDSS DR7 \citep{Abazajian2009}. We consider three different volume limited samples: $10^{9.8}\le \mstar < 10^{10.2}$, $10^{10.2}\le \mstar < 10^{10.6}$, $10^{10.6}\le \mstar < 10^{11.2}$, where the redshift limits of these samples are given in \cref{tab:sdss_samples}, and all stellar masses are quoted in units of $h^{-2}\msol$. Although the redshift ranges have been chosen to provide volume complete samples, we have restricted the most massive sample's upper redshift limit to a lower value than otherwise would be possible so that we do not need to account for redshift evolution of the stellar mass function in our SHAM models. We also subdivide these samples into star-forming and quenched galaxies using measurements of their specific star formation rate (SSFR), where we categorize galaxies with $\textrm{SSFR} < 10^{-11}\, \rm yr^{-1}$ as quenched. In order to avoid complications from the differences between the imaging used for target selection in the north galactic cap (NGC) and south galactic cap, we limit our analyses to the the NGC.

\begin{table}[]
    \centering
    \begin{tabular}{cccccc}
    \hline
    \hline 
        $\log_{10}M^{*}$ & $z_{\rm min}$ & $z_{\rm max}$ & $N_{\rm gal}$ & $N_{\rm Q}$ & $N_{\rm SF}$\\
    \hline
        10.6\textrm{ to } 11.2 & 0.026 & 0.106  & 29879 & 12932 & 16947\\
        10.2\textrm{ to } 10.6 & 0.026 & 0.106  & 76091 & 47502 & 28586\\
        9.8\textrm{ to } 10.2  & 0.026 & 0.067 & 21925 & 18201 & 3723\\
    \hline
    \end{tabular}
    \caption{SDSS sample definitions used throughout this work. The number of quenched and star forming galaxies does sum to the total number of galaxies in each stellar mass bin because some galaxies do not have good measurements of SSFR.}
    \label{tab:sdss_samples}
\end{table}
We measure three different clustering statistics for each sample: projected clustering, $w_p(r_p)$, and the monopole and quadrupole moments of the redshift-space clustering signal, $\xi_0(s)$ and $\xi_2(s)$. The projected correlation function is given by:

\begin{align}
    w_p(r_p) = 2\int_0^{\pi_{\rm max}} d\pi\, \xi(r_p, \pi),
\end{align}

where $\pi = \frac{\mathbf{s} \cdot \mathbf{l}}{|\mathbf{l}|}$, $r_p^2 = \mathbf{s} \cdot \mathbf{s} - \pi^2$, and $\mathbf{s} = \mathbf{s_1} - \mathbf{s_2}$, $\mathbf{s_1}$ and $\mathbf{s_2}$ are the redshift-space coordinates of two galaxies, and $\mathbf{l} = (\mathbf{s_1} + \mathbf{s_2}) / 2$ \citep{Davis&Peebles1983, Fisher1994}. We use $\pi_{\rm max} = 40\,\mpch$ for all of the $w_p(r_p)$ measurements presented in this work. We estimate $\xi(r_p, \pi)$ using the Landy-Szalay estimator \citep{Landy&Szalay1993}:

\begin{align}
    \xi(r_p, \pi) = \frac{DD - 2DR + RR}{RR},
\end{align}
where $DD$, $RR$, and $DR$ are the number of data-data, random-random and data-random pairs, normalized by the total number of pairs in each radial bin. All of the pair-counting done in this paper makes use of the \texttt{Corrfunc} library \citep{Sinha&Garrison2017}. We use 12 logarithmically spaced bins between $r_p=0.13-32.6\, \hmpc$, and 40 linearly spaced bins in $\pi$. We assume the \texttt{SMDPL} cosmology (see \cref{sec:sims}) to convert redshift to comoving LOS distance, always setting $h=1$.

The monopole and quadrupole moments of the anisotropic redshift-space clustering signal are given by:

\begin{align}
    \xi_{\ell}(s) = \int_{0}^{1} d\mu\, \xi(s,\mu)\mathcal{L}_{\ell}(\mu),
\end{align}
where $\mu = r_{p}/s$ is the cosine of the angle between the line-of-sight and $\mathbf{s}$. $\mathcal{L}_{\ell}$ is the $\ell$-th Legendre polynomial, with $\mathcal{L}_0 = 1$ and $\mathcal{L}_2=(3\mu^2 - 1)/2$. $\xi(s,\mu)$ is also estimated using the Landy-Szalay estimator. We use 12 logarithmically spaced bins between $s=0.13-32.6\, \hmpc$ and 40 linearly spaced bins between $0\le \mu < 1$. 

Due to the finite size of the SDSS fibers used to transmit light from the focal plane to the spectrographs, spectra of pairs of galaxies closer than the 55 arcsecond diameter of a fiber cannot both be observed. This angular size translates to $r_p = 0.12\, \hmpc$ at the highest redshift edge of the samples considered in this work. For $w_{p}(r_{p})$ we simply restrict our analysis to scales larger than this. For $\xi_{\ell}(s)$ we must be more careful, as these statistics can have contributions from $r_p<0.12\, \hmpc$ for $s>0.12\, \hmpc$. In order to remove this dependence, we first assign all galaxies with missing redshifts their "Nearest Neighbor" redshift as implemented in the NYU VAGC. Additionally, we exclude any $(s,\mu)$ bins that have any contribution from $r_p<0.12\, \hmpc$ before computing multipoles, i.e.
\begin{align}
    \hat{\xi}_{\ell}(s) = \int_{0}^{\mu_{\rm max}(s)} d\mu\, \xi(s,\mu)\mathcal{L}_{\ell}(\mu),
\end{align}
where $\mu_{\rm max} = r_{p,\rm max} / s$. \citet{Reid2014} showed that this approach is excellent at removing the bias imparted to $\xi_{0/2}(s)$ by fiber collisions at the expense of sensitivity to clustering at scales smaller than the fiber collision radius.

We estimate covariance matrices for all of the measurements presented here using a jackknife procedure. We use $N_{\rm SIDE}=8$ \healpix\ cells \citep{Gorski2005} as jackknife regions. We assign randoms to these \healpix\ cells, and exclude those cells that have fewer than $50\%$ of the average number density of randoms in them in order to ensure that our jackknife regions are of equal area. This results in 127 equal-area regions at the expense of removing 7253 galaxies that would have otherwise been included in our samples. We can then compute our covariance matrix as:

\begin{align}
\label{eq:jackknife}
    \textrm{Cov}(x_i, x_j) = \frac{N - 1}{N}\sum_{k=1}^{N} (x_{i,k} - \bar{x}_i) (x_{j,k} - \bar{x}_j),
\end{align}
where $x_i$ and $x_j$ are two elements of our data-vector, $x_{i,k}$ and $x_{j,k}$ are the same elements measured when leaving the $k$-th jackknife region out, and $\bar{x}_i$ and $\bar{x}_j$ are the means of those elements averaged over all $N=127$ jackknife measurements. We apply a Hartlap correction to our inverse covariance matrices \citep{Hartlap2007} in order ameliorate biases in our analysis due to noise in the covariance matrix. The largest data vector that we consider in this work, the combination of $w_p$ and \multipoles\ for all three stellar mass bins, has length $N_d=108$, so the Hartlap factor leads to a $87\%$ decrease in constraining power in this case. 

\section{Simulations}
\label{sec:sims}
In this work we make use of the Small Multi-Dark Planck (\texttt{SMDPL}) simulation \citep{Klypin2016}, an $N$-body simulation run using \texttt{L-Gadget2} \citep{Springel2005} with $3840^3$ particles in a $(400\, \hinv\mpc)^3$ volume and a force softening of $\epsilon_{\rm Plummer}=1.5\, \hkpc$, yielding a mass resolution of $9.63\times10^{7}\hinv\mpc$. Using a simulation with a volume of at least $(400 \hinv\mpc)^3$ is important when analyzing SDSS MGS clustering, otherwise sample variance from the simulation becomes a dominant contribution to SHAM parameter constraints \citep{Lehmann2017}. The simulation was initialized using the Zel'dovich approximation at $z=100$ and the best fit cosmological parameters from \citet{Planck2015}. Halo finding was performed using \texttt{Rockstar} \citep{Behroozi2013}, assuming a virial overdensity definition \citep{Bryan&Norman1998} and removing unbound particles from the halo mass estimates. Merger trees were generated for these halo catalogs using \texttt{Consistent Trees}, \citep{Behroozi2013b}, and orphan halos were simulated using \texttt{UniverseMachine} \citep{Behroozi2019}. Only the $z=0$ snapshot is used in this work.

\section{Models and Measurements from Simulations}
\label{sec:models}

Subhalo abundance matching is a technique that assigns galaxy stellar masses or luminosities to resolved dark matter (sub)halos by enforcing the relation:

\begin{equation}
\Phi(\mstar>x) = n(X_h>y),
\label{eq:abunmatch}
\end{equation}
where $\Phi(\mstar>x)$ is the cumulative number density of galaxies more massive than \mstar, and $n(X_h>y)$ is the cumulative number density of halos where some chosen halo property, $X_h$, often times referred to as the SHAM proxy, is greater than $y$. This implicitly defines a relation $\mstar(X_h)$ that can be used to assign galaxy stellar masses to halos as a function of $X_h$. 

We use the stellar mass function measured in \citet{Reddick2013} based on the data described in \cref{sec:obs}. Although this stellar mass function was measured over the redshift range $0.026 \le z < 0.067$, we have confirmed that using this stellar mass function accurately reproduces the number densities of our two most massive samples whose upper redshift limit extends to $z=0.106$. 

\Cref{eq:abunmatch} holds in the case that there is zero scatter in the relation between \mstar\ and $X_h$, but in any realistic scenario there \textit{is} scatter induced in this relation, both due to observational uncertainty in the measurement of \mstar\ and because of correlations between \mstar\ and variables in addition to $X_h$ that have been neglected in our model. We account for these sources of scatter by deconvolving a fiducial amount of scatter, $\sigma_{\log \mstar | X_{h}}$ (abbreviated as $\sigma_{\log \mstar}$ for the duration of this work), from $\Phi(x>\mstar)$, using \cref{eq:abunmatch} to determine $M_{r}(X_h)$, and then to adding the same amount of scatter back to the assigned values of \mstar. The deconvolution procedure assumes $p(\log \mstar|X_{h})$ to be log-normal, where $\sigma_{\log \mstar}$ is conventionally quoted in dex. \citet{Behroozi2010} describes this method in greater detail. $\sigma_{\log \mstar | X_{h}}$ is then left as a free parameter in the SHAM model to be fit to the data under consideration.

\subsection{SHAM Proxy}
In this work we use $v_{\alpha}$ as the proxy for \mstar\, i.e. $X_h=v_{\alpha}$ in \cref{eq:abunmatch}. $v_{\alpha}$ is given by:

\begin{align}\label{eq:valpha}
v_{\alpha} = v_{\rm vir}\left(\frac{v_{\rm max}}{v_{\rm vir}}\right)^{\alpha},
\end{align}
where $v_{\rm max}$ is the maximum circular velocity of the halo, $v_{\rm vir}=\frac{GM_{\rm vir}}{R_{\rm vir}}$, is the virial velocity of the halo, and $\alpha$ is a free parameter that governs the concentration dependence of the abundance matching proxy. This proxy is evaluated at the epoch at which the halo attains its peak mass, i.e. $v_{\rm max} = v_{\rm Mpeak}$ in \cref{eq:valpha}, and $v_{vir}$ is computed using $M_{\rm peak}$.

Using quantities evaluated at this epoch mitigates the effects of mergers on $v_{\rm max}$, which otherwise can cause large temporary spikes in $v_{\rm max}$ that likely do not contain information relevant to long term stellar mass evolution \citep{Behroozi2014,ChavesMontero2016}.

\subsection{Orphans}
\label{sec:orphan_model}
It has been shown that subhalos are susceptible to artificial disruption \citep{vandenbosch2018} even in high resolution \nbody\ simulations such as the one used in this work. Whether a subhalo artificially disrupts or not is strongly dependent on the orbit that the subhalo takes within its host halo, and as such can potentially impart biases on the shape of the clustering signals predicted by SHAM. 

The way that we account for this effect in our modeling is two-fold. First, we determine when a subhalo can no longer be tracked by our merger tree algorithm. After this point, we continue to evolve these disrupted subhalos forward in time, modeling their mass and velocity evolution as well as their orbits within their host halo semi-analytically in the manner described in Appendix B2 of \citet{Behroozi2018}. This procedure gives us a catalog of disrupted subhalos, also known as orphan subhalos, in addition to all the subhalos that can still be identified and tracked in the standard manner. 

A problem remains that many of the disrupted subhalos that we continue to track are actually disrupted in a physical manner, for example through major mergers. We must decide which of these orphan subhalos may actually still host galaxies. There is a growing literature that attempts to address this problem as a function of properties of the subhalo itself and its orbital parameters \citep{Ogiya2019, Jiang2020}. In this work, we parameterize the probability that an orphan subhalo physically disrupts, i.e. that it is not available to be populated with a galaxy, as a function of the maximum velocity of the subhalo at its peak mass, \vmpeak:

\begin{align}
    P(\textrm{disrupt}) = \Theta(T_{\rm disr}(\vmpeak) - v_{\rm now}/\vmpeak))\, ,
\end{align}
where $\Theta$ is the Heavyside step-function, and 

\begin{align}
    T_{\rm disr}(\vmpeak) &= T_{\rm disr, low} + (T_{\rm disr, high} - T_{\rm disr, low}) \times \nonumber \\ 
    & 0.5 + 0.5 \textrm{erf}\, \left ( \frac{\log_{10}(\vmpeak) - v_{\rm disr,mean}}{2\sigma_{\rm disr}} \right)\, ,
\end{align}
where $T_{\rm disr, low}$, and $T_{\rm disr, high}$ are the asymptotes of $T_{\rm disr}(\vmpeak)$ at low and high \vmpeak\ respectively, $\sigma_{\rm disr}$ determines the gradient of $T_{\rm disr}(\vmpeak)$ as a function of \vmpeak\ and $v_{\rm disr,mean}$ sets the \vmpeak\ at which the slope of $T_{\rm disr}(\vmpeak)$ is greatest. We also investigated a number of other forms for $T_{\rm disr}$, including dependence on host \vmpeak\ and a joint dependence on host and subhalo \vmpeak, none of which improved our ability to fit all of the stellar mass bins in our data simultaneously, as discussed in \cref{sec:disruption}.

It is also possible that baryonic effects disrupt subhalos as a function central galaxy morphology (e.g. the presence of a disk), star formation rate or super-massive black hole activity. In this work we have assumed that all subhalos that are resolved at $z=0$ in our simulation can potentially host galaxies, and thus baryonic effects cannot disrupt resolved subhalos. This seems a reasonable assumption given the stellar mass ranges considered here, especially as there is only a very small population of halos that are significantly stripped but still resolved in our simulations.

\subsection{Velocity Bias}
\label{sec:velocity_bias_model}
As an additional extension of our model we consider the possibility that the velocity distributions of central and satellite galaxies differ from the velocity distributions of the (sub)halos in our simulations. We introduce two parameters to govern these potential deviations, $\alpha_{c}$ and $\alpha_s$, which determine central and satellite velocity bias respectively. 

We include central velocity bias by assigning central galaxies LOS velocities, $v_c$, with respect to the LOS host halo center-of-mass velocity, $v_h$, drawing from the distribution:
\begin{align}
    p(v_{c} - v_{h}) = \frac{1}{\sqrt{2\pi} \sigma_c} \exp{\left(-\frac{(v_{c} - v_{h})^2}{2\sigma_c^2}\right)}\, ,
\end{align}
where $\sigma_c=\alpha_c \sigma_h/\sqrt{3}$, $\sigma_h$ is the three-dimensional host halo velocity dispersion as measured by \rockstar, and the factor of $\sqrt{3}$ comes from the conversion between three-dimensional velocity dispersion and LOS velocity. 

Satellite galaxies, i.e. galaxies that have been assigned to subhalos, are treated separately. In the absence of satellite velocity bias, satellite galaxies are assigned the velocities of the subhalos that host them. Our implementation of satellite velocity bias simply re-scales the satellites velocities in the host-halo frame of reference:
\begin{align}
v_{s} = \alpha_s(v_{\rm sub} - v_{h}) + v_h\, ,
\end{align}
where $v_{s}$ is the satellite LOS velocity, and $v_{\rm sub}$ is the LOS velocity of the subhalo that the satellite galaxy is assigned to.

\subsection{Conditional Abundance Matching}
\label{sec:cam_model}
In order to assign galaxy SSFRs to simulated galaxies, we adopt a conditional abundance matching (CAM) framework. CAM posits that at fixed stellar mass, SSFR is monotonically related to a second halo property, $Y_{\rm halo}$, i.e.
\begin{align}
\label{eq:cam}
    F(\textrm{SSFR} | \mstar) = P(<\textrm{SSFR} | \mstar) = P(<Y_{\rm halo} | \mstar)
\end{align}
where $F(\rm SSFR | \mstar)$ is the cumulative distribution of galaxy SSFR at fixed \mstar. No functional form is assumed for $P(<\rm SSFR | \mstar)$ or $P(<Y_{\rm halo} | \mstar)$, rather, they are measured directly from the data and the simulations respectively. 

As a generalization of CAM, we allow for an imperfect correlation between $Y_{\rm halo}$ and SSFR. This is accomplished by enforcing
\begin{align}
\label{eq:noisycam}
    P(< \textrm{SSFR} | \mstar) = P(<\widetilde{Y}_{\rm halo} | \mstar)
\end{align}
where
\begin{align*}
    P\left(\sqrt{2} \, S(2 \widetilde{\mathcal{R}} - 1) \Bigm\lvert \,
    \sqrt{2} S(2 \mathcal{R} - 1)\right) \\ 
    \sim \mathcal{N} \left (\sqrt{2} S(2 \mathcal{R} - 1)), r\right) \nonumber\, ,
\end{align*}
with $S(x) = \textrm{erf}^{-1}(x)$, $\widetilde{\mathcal{R}} = \textrm{Rank}(\widetilde{Y}_{\rm halo})$ and 
$\mathcal{R} = \textrm{Rank}(Y_{\rm halo}) \in [0,1]$. This ensures that $\widetilde{Y}_{\rm halo}$ is a noisy version of $Y_{\rm halo}$, where the Pearson correlation coefficient between ${\rm Rank}(Y_{\rm halo})$ and ${\rm Rank}(\widetilde{Y}_{\rm halo})$ is set to $r$.

We use a "bin-free" version of CAM as implemented in \texttt{halotools} \citep{Hearin2016}, where $P(<Y_{\rm halo} | \mstar)$ and $P(<\textrm{SSFR} | \mstar)$ are determined in sliding windows around the $M^{*}$ of each simulated halo and galaxy in the data respectively. This allows us to avoid the discreteness effects imparted by wide bins in $\mstar$ evident in earlier implementations of CAM.

In a similar way to \cref{eq:abunmatch}, this defines a relation $P(\textrm{SSFR}|Y_{\rm halo},\mstar)$. In this work we consider two different quantities for $Y_{\rm halo}$. The first, which we refer to as $\Delta v_{\rm max}$, is a measure of recent accretion of matter onto subhalos defined as
\begin{align}
\label{eq:dvmax}
    \Delta v_{\rm max}(a) = \frac{v_{\rm max}(a)}{v_{\rm max}(\textrm{min}[a_{\rm Mpeak},a_{\rm dyn}])}\, ,
\end{align}
where $a_{\rm Mpeak}$ is the scale factor at which the halo attains its peak mass, and $a_{\rm dyn}$ is the scale factor one dynamical time before $a$, where a dynamical time is given by $t_{\rm dyn}=\left(\frac{4}{3}\pi G \rho_{\rm vir}\right)^{-1/2}$. \dvmax\ is also used as a proxy for SFR in \texttt{UniverseMachine} \citep{Behroozi2019}, albeit in a significantly more intricate manner.

The second proxy that we make use of is $R_{\rm h}$. \rhalo\ is defined as the distance between a given galaxy and the closest host halo with mass greater than a specified threshold, \mcut, where \mcut\ is left as a free parameter that is fit to data. This proxy is motivated by claims that quenching has a strong dependence on proximity to massive halos, and that there is a particular mass scale for this quenching \citep{Peng2010,Behroozi2013,Zu2015}.

\subsection{Measurements in Simulations}
\label{sec:sim_measurements}
We take a slightly different approach to RSD measurements in our simulation than what we use on the SDSS data. In particular, the periodic nature of our simulations allows measurements to be made without the use of random points. This is important in order to improve the accuracy of the emulators we build in this work, which make use of small scale clustering measurements. Otherwise, a very large number of random points would be required to remove the contribution of their finite sampling from our measurements. Because of this difference between the measurements in our simulations, and those made in the data, the angular and redshift window functions of the data are not accounted for in our model predictions. Nevertheless, these are largely accounted for in our jackknife covariance matrix measured from the SDSS data.

For each stellar mass bin in \cref{tab:sdss_samples}, we select an analogous sample in our simulations by cutting on the stellar masses that have been assigned via our SHAM model. We produce redshift-space coordinates for each galaxy from real-space positions by modifying the LOS coordinate as follows:

\begin{align}
   x_{z,{\rm rsd}} = x_{z, {\rm comov}} + \frac{v_{z}}{a H(a)},
\end{align}
where $x_{z,{\rm rsd}}$ and $x_{z,{\rm cos}}$ are the redshift-space and comoving line-of-sight coordinates and $v_z$ is the line-of-sight velocity in $\rm km\,s^{-1}$. Periodic boundary conditions are applied after this transformation. We can then use the natural estimator:
\begin{align}
    \xi = \frac{DD}{RR} - 1
\end{align}
for $\xi(r_p, \pi)$ and $\xi(s, \mu)$, where $DD$ and $RR$ are again data-data and random-random pair counts normalized by the number of pairs in each bin. The $RR$ term can now be expressed analytically, given our periodic boundary conditions. $w_p(r_p)$, $\hat{\xi}_0$ and $\hat{\xi}_2$ are then calculated in the same way as described in \cref{sec:obs}. We measure $w_p(r_p)$, $\hat{\xi}_0$ and $\hat{\xi}_2$ three times, once using each of the $x$, $y$, and $z$ axes of our simulation as the line-of-sight, using the average over the three lines of sight to build our emulators.

For the purposes of estimating covariance matrices for our simulation measurements, which is necessary due to the non-negligible contribution of sample variance from our simulations to the total error in our analysis, we again use a jackknife procedure. Here we make use of 125 equal volume sub-boxes as our jackknife regions rather than the regions used for measurements on SDSS. Our covariance matrix is again estimated from our jackknife measurements using \cref{eq:jackknife}. We compute this jackknife covariance matrix at the joint best fit parameter values for $w_p$, $\hat{\xi}_0$ and $\hat{\xi}_2$ for the \dvmax\ CAM model discussed in \cref{sec:dvmax_cam}.

We must account for one additional source of uncertainty in our analysis: the stochasticity imparted on our measurements due to our use of Monte Carlo draws from random variables in our SHAM and CAM implementations. We do this by repopulating our simulations 10 times at each point in SHAM and CAM model space and recomputing $w_p$, $\hat{\xi}_0$ and $\hat{\xi}_2$  for each re-population. The measurements presented in the following sections are the mean of these 10 re-populations. We neglect any residual contribution of this stochasticity in our error budget. 

\section{SHAM Results}
\label{sec:base_sham}
Instead of directly populating our simulations at each point in parameter space in order to make clustering predictions, we construct surrogate models for our SHAM and CAM models and use these to predict observables as a function of our model parameters. \Cref{app:emu} describes the surrogate modeling framework we use to perform all of our parameter estimation in more detail.

All posterior parameter distributions and evidences in this work are derived using the nested sampling algorithm \texttt{dynesty} \citep{Speagle2019}. We use a convergence criterion of $\Delta \log Z < 0.01$ for all analyses, where $Z$ is the evidence. All analyses assume a Gaussian likelihood, where the covariance used is the combination of the jackknife covariance matrix estimated from the data and the jackknife covariance matrix from the simulations used as an approximation for the error on our surrogate models. We assume flat priors on all parameters, whose edges are listed in \cref{tab:params}. 

\begin{table*}
\caption{Parameters and priors}
\begin{center}
\begin{tabular}{ c  c  c }
\hline
\hline
Parameter & Prior & Analysis Configuration\\  
\hline 
\multicolumn{3}{c}{{\bf Base SHAM}} \\
$\sigma_{\log_{10}M^*}$  &  flat (0.0, 0.8) & \cref{sec:base_sham} \\ 
$\alpha$  &  flat (0.0, 1.0) & \cref{sec:base_sham} \\
\hline
\multicolumn{3}{c}{{\bf Sub-halo Disruption}} \\
$T_{\rm disr, high}$ & flat (0.2, 1.4) & \cref{sec:disruption} \\ 
$T_{\rm disr, low}$ &  flat (0.2, 1.4) & \cref{sec:disruption} \\ 
$v_{\rm disr, mean}$ & flat (1.9, 3.3) & \cref{sec:disruption} \\ 
$\sigma_{\rm disr}$ &  flat (0.1, 4) & \cref{sec:disruption} \\ 

\hline
\multicolumn{3}{c}{{\bf Velocity Bias}} \\
$\alpha_{c}$ & flat (0.0, 0.5) & \cref{sec:velocity_bias} \\ 
$\alpha_{s}$ &  flat (0.5, 1.5) & \cref{sec:velocity_bias} \\
\hline
\multicolumn{3}{c}{{\bf \dvmax\ CAM}} \\
$r_{\dvmax}$ & flat (0.0, 1.0) & \cref{sec:dvmax_cam} \\ 

\hline 
\multicolumn{3}{c}{{\bf \rhalo\ CAM}} \\
$r_{\rhalo}$ & flat (0.0, 1.0) & \cref{sec:rhalo_cam} \\ 
$\log_{10}M_{\rm cut}$ & flat (13, 15) & \cref{sec:rhalo_cam} \\ 
\hline
\end{tabular}
\end{center}
\label{tab:params}
\end{table*}

\subsection{Baseline SHAM}
We now study the behavior of our fiducial SHAM model including as parameters only the scatter in stellar mass at fixed SHAM proxy, $\sigma_{\log M^{*}}$, and the parameter controlling the concentration dependence of the SHAM proxy, $\alpha$. 

\begin{figure*}
\centering
      \includegraphics[width=\textwidth]{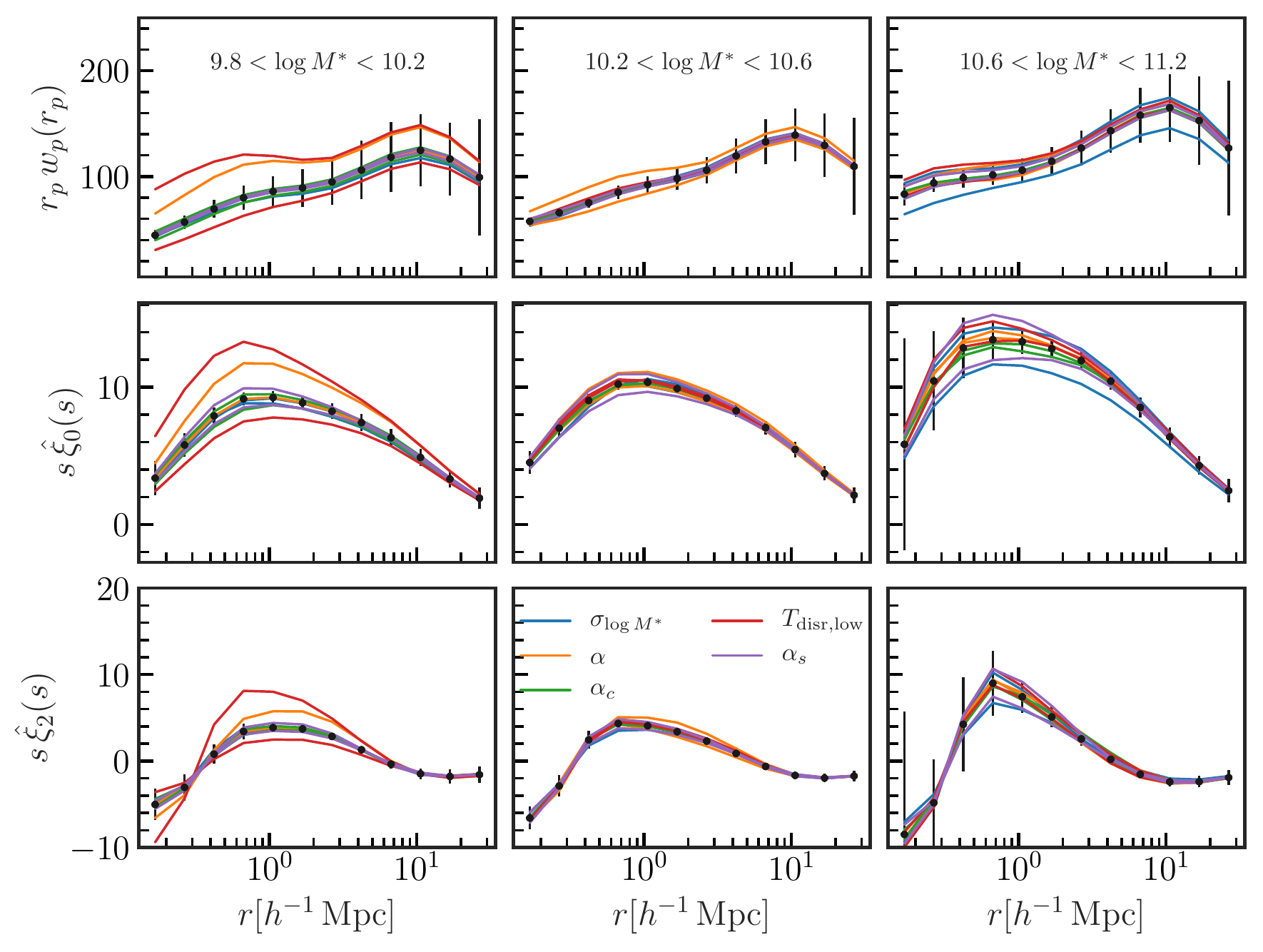}
  \caption{Effect of varying parameters of our SHAM model by $\pm 2\sigma$ around their best fit values for each stellar mass bin. The solid points with error bars are the best fit model predictions, while the colored lines are the model predictions varying each parameter one by one. We see that the most important parameter for each stellar mass bin is different, with $\sigma_{\log M^*}$, $\alpha$, and $T_{\rm disr,low}$ causing the largest variations in the most, second most, and least massive stellar mass bins respectively. Satellite velocity bias is important for \multipoles\ for all masses, having a small residual impact on $w_p$ because we do not project over infinitely long distances along the LOS. $\alpha$, and $T_{\rm disr,low}$ have similar impacts on large scales, but differ for $r<2 \hmpc$, allowing some ability to constrain $\alpha$ in the presence of subhalo disruption systematics.}
  \label{fig:derivatives}
\end{figure*}

The effect that these parameters have on projected and redshift-space clustering can be seen in \cref{fig:derivatives}, where we show the effect of varying $\sigma_{\log M^{*}}$ and $\alpha$ (among other parameters to be discussed later) by $2\sigma$ around their best fit values for each stellar mass bin. It is clear that $\sigma_{\log M^{*}}$ has the greatest impact on the most massive bin, affecting $w_p$ and \multipoles\ at similar levels. This is expected, as the effect that variations in $\sigma_{\log M^{*}}$ have on clustering is highly dependent on the slope of the stellar mass function, and the bias--halo-mass relation $b(M)$. This is because increasing $\sigma_{\log M^{*}}$ preferentially brings galaxies hosted by lower mass halos into a given stellar mass selection. This effect is stronger the steeper the slope of the stellar mass function. As galaxies hosted by halos of lower mass scatter into the selection, the large scale bias of that sample is reduced because of the positive slope of $b(M)$. Thus, changes in $\sigma_{\log M^{*}}$ have the greatest effect where the slope of $b(M)$ and stellar mass function are simultaneously large, which occurs at high stellar masses.

Variations in $\alpha$ become quite important for the two less massive bins, where there is more scatter in $\vmpeak-v_{\rm vir}$ relation \citep{Lehmann2017}. Increasing $\alpha$ ranks halos with larger concentrations higher at fixed \mpeak, thus preferentially selecting satellite galaxies, which generally form earlier and are thus more concentrated on average than host halos of the same \mpeak. As such, increasing  $\alpha$ boosts the satellite fraction $f_{\rm sat}$ of our samples, preferentially increasing both the one halo term of our clustering signals, which scales as the number of satellite galaxies squared, and the large scale bias of our samples, as satellites preferentially reside in more massive and highly biased halos. Increasing $\alpha$ also preferentially selects more concentrated central halos at fixed \mpeak, again boosting the large scale bias due to the secondary dependence of halo bias on concentration, which is a positively sloped relation at this halo mass scale \citep{Wechsler2006,Mao2017}. We will revisit the effects of these parameters on our clustering statistics when discussing model extensions in \cref{sec:velocity_bias,sec:disruption}, but we note here that additional SHAM parameters such as artificial subhalo disruption and velocity bias severely hamper our ability to constrain $\alpha$.

\Cref{tab:sham_results} lists the reduced chi-squared values of the best fit models for a number of different data vector and galaxy sample combinations. For the two most massive galaxy samples ($10.2\le \log_{10}M^{*} < 10.6$ and $10.6\le \log_{10}M^{*} < 11.2$, combined referred to as "Top two") our model obtains a good fit to $w_p$, $\hat{\xi}_0$ and $\hat{\xi}_2$ simultaneously. All three clustering measurements, $w_p$ and \multipoles, show similar reduced chi-squared values, suggesting that there is not significant tension in the model when trying to fit all three statistics simultaneously. Nevertheless, we show in \cref{sec:disruption,sec:velocity_bias} that some model extensions can improve these fits.

We also see in \cref{tab:sham_results} that the goodness of fit for all clustering statistics is significantly worse for the $9.8\le \log_{10} M^{*} < 10.2$ sample than for the more massive samples. This is particularly true for the redshift-space clustering statistics, and can also be seen in the relatively larger values for reduced chi-squared seen in the joint fit to all three samples. In the following sections, we will discuss further extensions to our fiducial SHAM model that improve our ability to fit this least massive galaxy sample, although no extension does a good job of simultaneously fitting all three samples.

\begin{figure}
\centering
      \includegraphics[width=\linewidth]{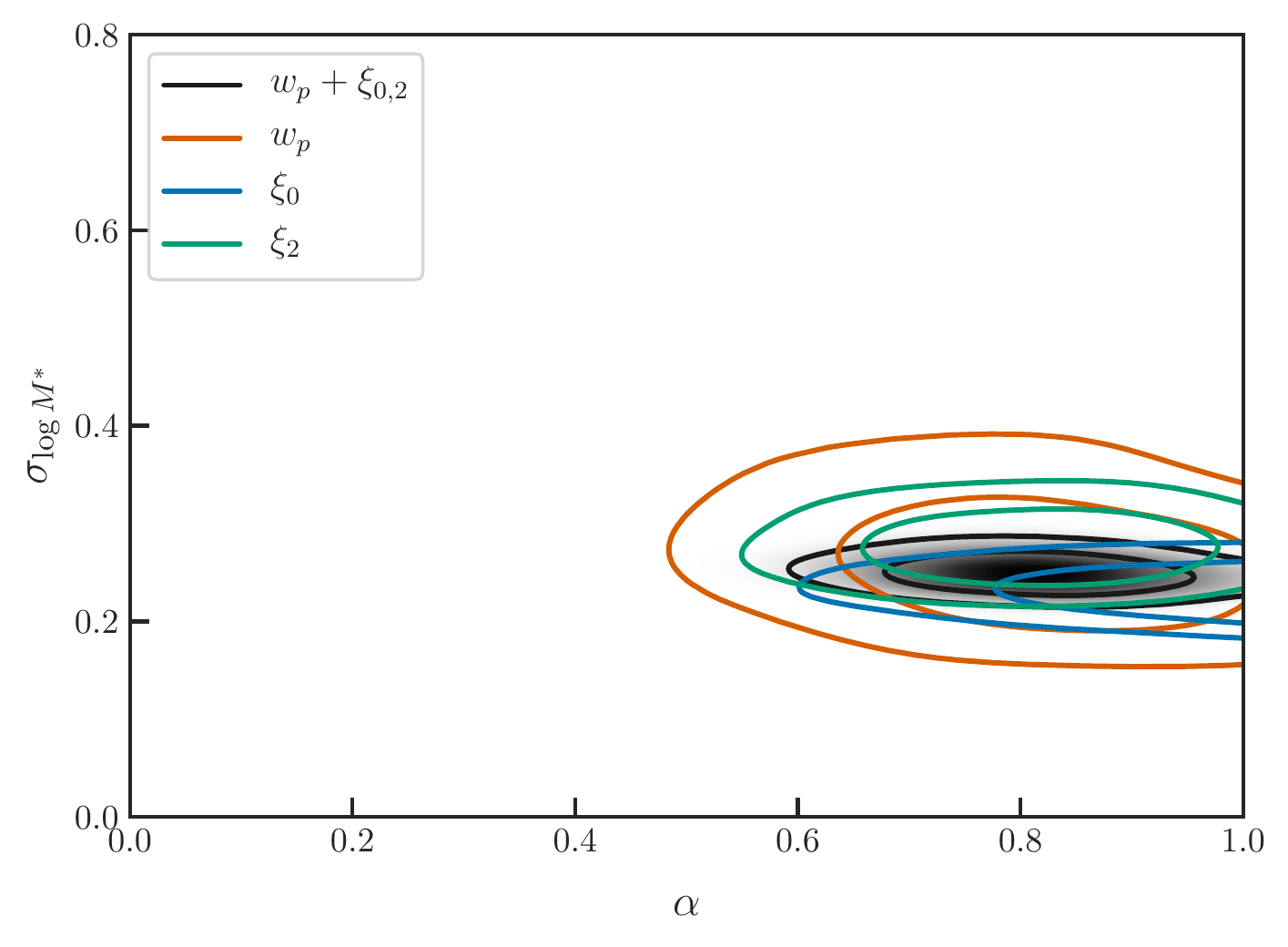}
  \caption{Constraints on the parameters of the fiducial SHAM model, $\sigma_{\log M^{*}}$, log-normal scatter in $M^{*}$ at fixed $v_{\alpha}$, and $\alpha$, the parameter that controls the concentration dependence of the SHAM proxy, fit to the two most massive galaxy samples. Various contours are constraints from different clustering measurements, all of which are statistically consistent. The combined result rules out $M_{\rm peak}$ ($\alpha=0$) SHAM at many sigma, and $v_{Mpeak}$ SHAM ($\alpha=1$)at 1.4 sigma.}
  \label{fig:fid_sham_comp}
\end{figure}

\Cref{fig:fid_sham_comp} shows constraints on the fiducial SHAM model parameters when fit to statistics measured from the two most massive galaxy samples. We see that all three statistics prefer consistent values for the fiducial SHAM parameters. The parameter constraints from each statistic have very similar degeneracies, with $\sigma_{\log_{10}M^*}$ and $\alpha$ showing little correlation. $\xi_{0}$ and $\xi_{2}$ are significantly more constraining than $w_{p}$, mostly due to the relatively larger signal to noise of the multipole measurements. 

\begin{figure*}
\centering
      \includegraphics[width=\textwidth]{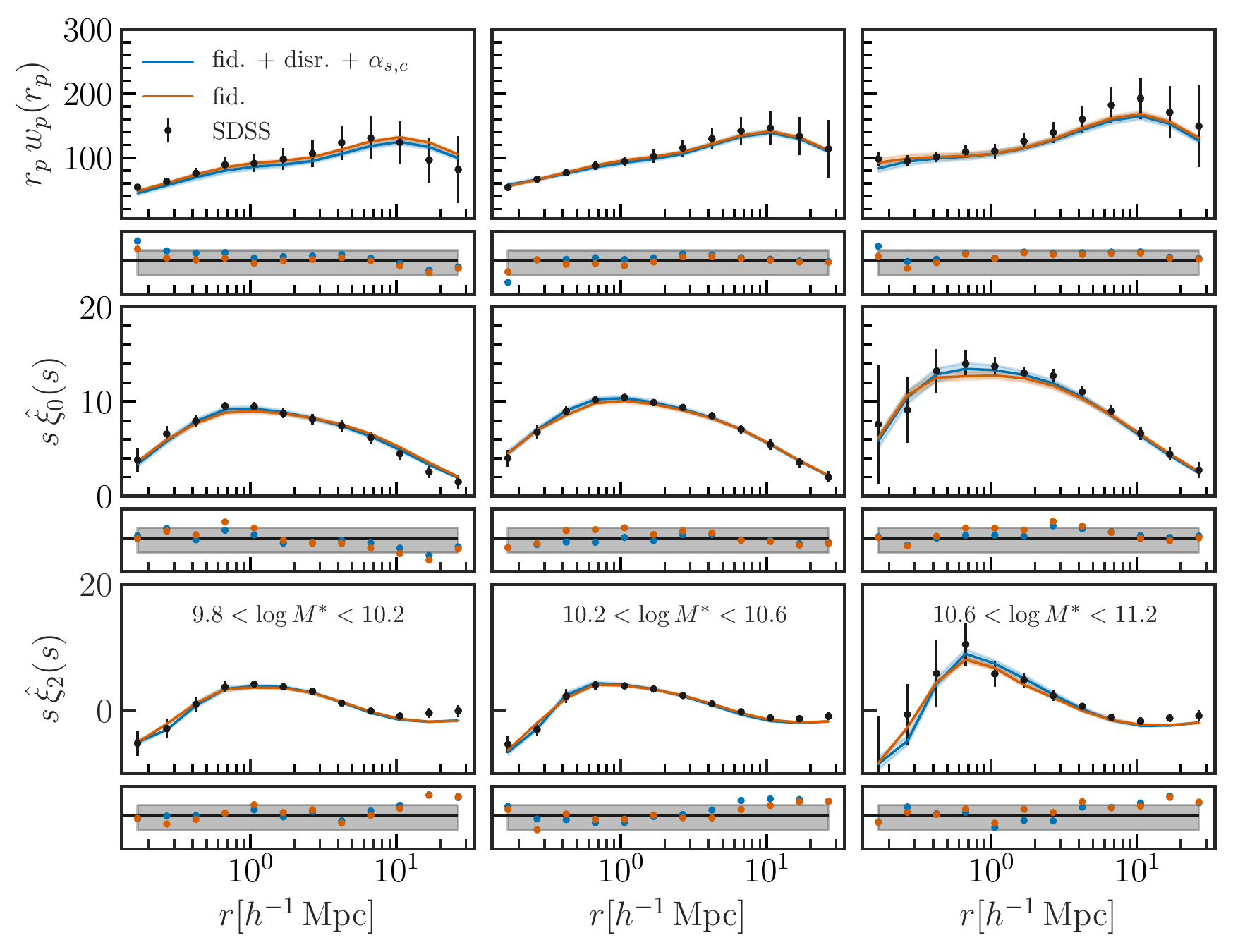} 
  \caption{Comparison of the best fit projected correlation function, $w_p(r_p)$ , monopole $\xi_{0}(s)$ and quadrupole $\xi_{2}(s)$ models (lines) models the SDSS measurements (points) in three bins of stellar mass, as listed in the bottom row. Rows alternate between clustering measurements, and fractional residuals of our model from the measured data. Gray bars in the fractional residual panels represent the $1\sigma$ errors. Error bars on the points include both sample variance from the data and from our simulation. Shaded regions around the solid lines represent the $1-\sigma$ posteriors of our model. The blue lines are the best fit model including artificial subhalo disruption and velocity bias as discussed in \cref{sec:disruption,sec:velocity_bias}, while the orange line is our fiducial model, discussed in \cref{sec:base_sham}. Velocity bias is important in order to fit the most massive sample, and orphan galaxies are preferred by the two lower mass samples.}
  \label{fig:sham_comparison}
\end{figure*}

We also show the best fit model to all clustering statistics in each stellar mass bin in \cref{fig:sham_comparison}. This figure conveys the same impression as the goodness-of-fit statistics in \cref{tab:sham_results}, namely that our model is a good fit to the two most massive galaxy samples and that there are more significant residuals in the least massive sample.

Our ability to fit the more massive samples runs contrary to the results presented in \citet{Guo2016}, who used the same simulation to show that a SHAM model accounting for only scatter in absolute magnitude at fixed SHAM proxy cannot simultaneously fit projected and redshift-space clustering in the SDSS MGS. This discrepancy can be explained by considering the constraints that we obtain on $\alpha$, which governs the behavior of the mass proxy used in our abundance matching models. In \cref{fig:sham_comparison} we see that our data rule out $\alpha=0$, which corresponds to $M_{\rm peak}$ abundance matching, at high significance. $M_{\rm peak}$ is approximately the same as the $M_{\rm acc}$ abundance matching model presented in \citet{Guo2016}, and so we see that when using a similar SHAM proxy we obtain similar results. Our constraints rule out $\alpha=1$, corresponding to $\vmpeak$ at $1.4\sigma$ confidence. $\vmpeak$ abundance matching is similar, but not identical, to the $v_{\rm acc}$ and $v_{\rm peak}$ abundance matching models that \citet{Guo2016} also rule out, although \citet{Guo2016} rules them out at much higher significance. This discrepancy may be a result of slight differences between $\vmpeak$, $v_{\rm acc}$ and $v_{\rm peak}$, and may also be a result of different fitting procedures and galaxy samples. \citet{Guo2016} also use a different method to correct for fiber collisions than the one used here, allowing them to use smaller scale clustering measurements, which could also account for some of the differences between the findings presented here and those presented in their work.

Nevertheless, we see that allowing for a continuous degree of freedom governing the concentration dependence of our abundance matching proxy is important for obtaining good fits to all clustering statistics simultaneously. We have tested that this still holds when selecting galaxies using absolute magnitude in \cref{app:absmag}, as was done in \citet{Guo2016}. Although we obtain improved fits with respect to those presented in \citet{Guo2016}, we shall show in \cref{sec:disruption,sec:velocity_bias} that extensions to the baseline SHAM model presented here are still preferred by the data in some cases.

\begin{table*}
\caption{Fiducial SHAM Reduced Chi-Squared Values}
\begin{center}
\begin{tabular}{ c  c  c  c  c  c }
\hline
\hline
 & $9.8\le \log_{10}M^{*} < 10.2$ & $10.2\le \log_{10}M^{*} < 10.6$ &
 $10.6\le \log_{10}M^{*}$ < 11.2 & Top Two & All\\
\hline 
$w_p + \hat{\xi}_{0,2}$ & 1.14 & 0.73 & 0.67 & 0.70 & 1.54 \\
$w_p$ & 0.93 & 0.51 & 0.27 & 0.40 & 0.62\\
$\hat{\xi}_{0}$ & 1.38 & 0.95 & 0.34 & 0.54 & 0.80\\
$\hat{\xi}_{2}$ & 0.89 & 0.44 & 0.66 & 0.44 & 0.78\\
\hline 
\end{tabular}
\end{center}
\label{tab:sham_results}
\end{table*}

\begin{table*}
\caption{SHAM Bayes Factors}
\begin{center}
\begin{tabular}{ c  c  c  c  c  c }
\hline
\hline
 & $9.8\le \log_{10}M^{*} < 10.2$ & $10.2\le \log_{10}M^{*} < 10.6$ &
 $10.6\le \log_{10}M^{*}$ < 11.2 & Top Two & All\\
\hline 
Fid. + orphans & $\mathbf{2.8 \pm 0.3}$ & $\mathbf{2.1 \pm 0.3}$ & $0.37 \pm 0.06$ & $0.40 \pm 0.09$ & $0.0017 \pm 0.0004$ \\
Fid. + $\alpha_{c,s}$       & $\mathbf{3.1 \pm 0.4}$ & $0.4 \pm 0.4$ & $\mathbf{6.80 \pm 0.06}$ & $0.4 \pm 0.1$ & $0.046 \pm 0.0004$\\
Fid. + orphans + $\alpha_{c,s}$ & $\mathbf{6.2 \pm 0.4}$ & $0.4 \pm 0.4$ & $\mathbf{1.84 \pm 0.07}$ & $0.9 \pm 0.1$ & $\mathbf{1 \pm 0.0005}$ \\
\hline 
\end{tabular}
\end{center}
\label{tab:sham_bayes_factor}
\end{table*}

\subsection{Orphan galaxies}
\label{sec:disruption}
We now examine physically motivated extensions to our fiducial SHAM model to see how they impact our parameter constraints and goodness-of-fit statistics. We consider two different model extensions: orphan galaxies and velocity bias.  

First, we consider a SHAM extension that corrects for potential artificial subhalo disruption in SMDPL. We will often refer to this extension as an "orphan" model. Artificial subhalo disruption has received attention in the recent literature, with \cite{vandenbosch2018} pointing out that artificial disruption is commonplace in cosmological $N$-body simulations at virtually all resolutions, even for subhalos that are resolved with hundreds of thousands of particles.

The model that we use to correct for this potential simulation artifact is presented in \cref{sec:orphan_model}. Orphan subhalos are tracked after disruption until their current maximum circular velocity, $v_{\rm now}$, falls below a fraction, $T_{\rm disr}$, of their \vmpeak\ value. The four parameters of the orphan model govern the \vmpeak\ dependence of $T_{\rm disr}$, although only one of them, the parameter controlling the low \vmpeak\ asymptote of $T_{\rm disr}$, $T_{\rm disr, low}$, has a large effect on the SHAM model predictions. As we decrease $T_{\rm disr, low}$, we bring more subhalos into our sample, thus increasing $f_{\rm sat}$, again boosting the one and two-halo terms of our clustering signals. Changes in $T_{\rm disr, low}$ almost exlusively change $f_{\rm sat}$, and thus give the dependence of our clustering signals slightly different scale and environmental dependence than changes in $\alpha$ and $\sigma_{\log M^*}$. This can be seen in \cref{fig:derivatives}, where $T_{\rm disr, low}$, is largely degenerate with $\alpha$ and $\sigma_{\log_{10}M^*}$ on large scales, but on small scales has a different scale dependence especially in $\hat{\xi}_{0,2}$, where the effect of $f_{\rm sat}$ has an additional effect in boosting the amplitude of the finger-of-god, allowing for constraints on subhalo disruption that are not entirely degenerate with $\alpha$ and $\sigma_{\log M^*}$.

In order to perform model comparisons, we make use of Bayes factors:
\begin{align}
    \mathcal{R} = \frac{P(\mathbf{d} | M_{\rm ext})}{P(\mathbf{d} | M_{\rm fid})} = \frac{Z_{\rm ext}}{Z_{\rm fid}}\, ,
\end{align}
i.e. the ratio of the Bayesian evidence for the data given an extended model, $P(\mathbf{d} | M_{\rm ext})$ to that obtained using our fiducial model, $P(\mathbf{d} | M_{\rm fid})$. If the ratio is larger than unity, then the extended model is preferred over the fiducial model, and if it is less than unity then the data prefer the fiducial model. In \cref{tab:sham_bayes_factor}, we list the Bayes factors for each of our extended models computed when fitting the clustering measurements of each stellar mass bin, the two most massive mass bins, and all mass bins simultaneously. For the most massive bin we find $\mathcal{R}<1$, but for the two less massive bins the SDSS data do prefer the presence of orphan galaxies.

It is not surprising that the less massive bins considered here require orphan galaxies while the most massive bin does not. The effect of artificial subhalo disruption has been shown to have a halo mass dependence at fixed simulation mass resolution, with less massive and more poorly resolved subhalos disrupting sooner than their counterparts that are resolved with more particles \citep{Ogiya2019}. 

\begin{figure}
\centering
      \includegraphics[width=\columnwidth]{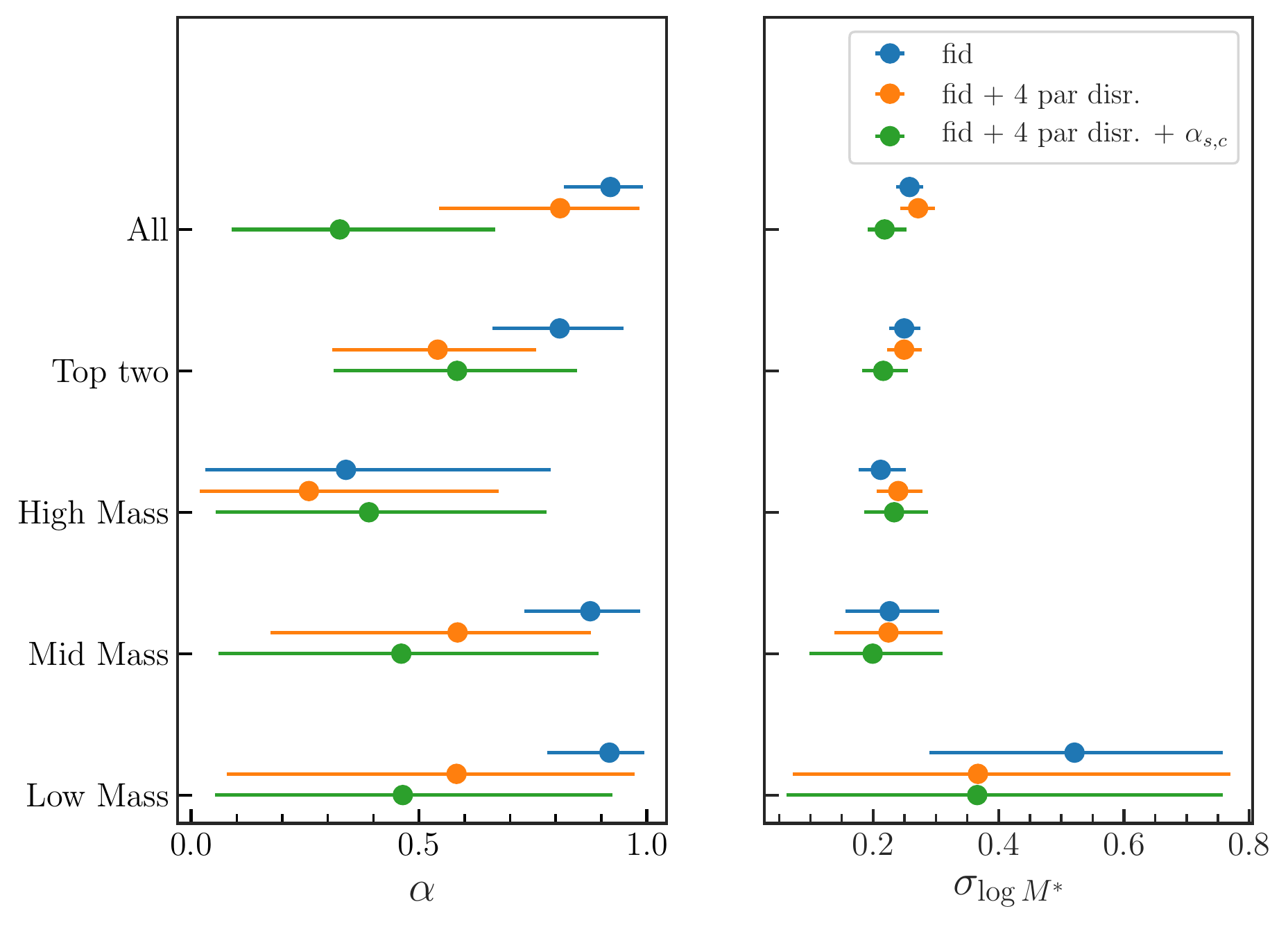} 
  \caption{{(\it Right)} Constraints on $\sigma_{\log_{10}M^*}$ for all three stellar mass bins individually, the two most massive bins analyzed simultaneously ("Top two") and all three bins analyzed simultaneously ("All"). Points are the 1 dimensional posterior means, and error bars depict $95\%$ confidence intervals. Different colored points represent different model choices, including our fiducial two parameter SHAM model (blue), an extension that allows for orphan galaxies (orange) and an extension allowing for both orphans and velocity bias (green). The constraints on $\sigma_{\log_{10}M^*}$ are not significantly affected by varying these modeling choices, except for the least massive bin where inclusion of orphans significantly degrades the ability to constrain $\sigma_{\log_{10}M^*}$. ({\it Left}) Same as right panel, but showing constraints on $\alpha$. Allowing for orphan galaxies significantly degrades the constraints in $\alpha$ for the two less massive galaxy samples. This indicates that previous preferences for \vmpeak\ abundance matching over $M_{\rm peak}$ abundance matching in the literature may have been driven by artificial subhalo disruption.}
  \label{fig:fid_model_comp}
\end{figure}

The larger impact of orphans on less massive samples is also observable in the effect that including orphan galaxies has on our $\sigma_{\log_{10}M^*}$ and $\alpha$ constraints, shown in \cref{fig:fid_model_comp}. The constraints on these parameters for the most massive galaxy sample are significantly less affected by including orphans in our model than the constraints from the less massive bins. For the least massive sample both $\alpha$ and $\sigma_{\log_{10}M^*}$ become almost entirely unconstrained once orphans are included, and for the second most massive bin, $\alpha$ becomes largely unconstrained. 

The changes in the constraints on $\alpha$ for the less massive bins are particularly interesting. The fiducial model favors $\alpha$ consistent with 1 for these bins, while including orphan galaxies removes nearly all constraining power on $\alpha$ suggesting that previous results favoring \vmpeak\ abundance matching over $M_{\rm peak}$ were largely a result of those models' lack of orphan galaxies, similar to the claims made in \citet{campbell2016}.

\begin{figure}
\centering
      \includegraphics[width=\columnwidth]{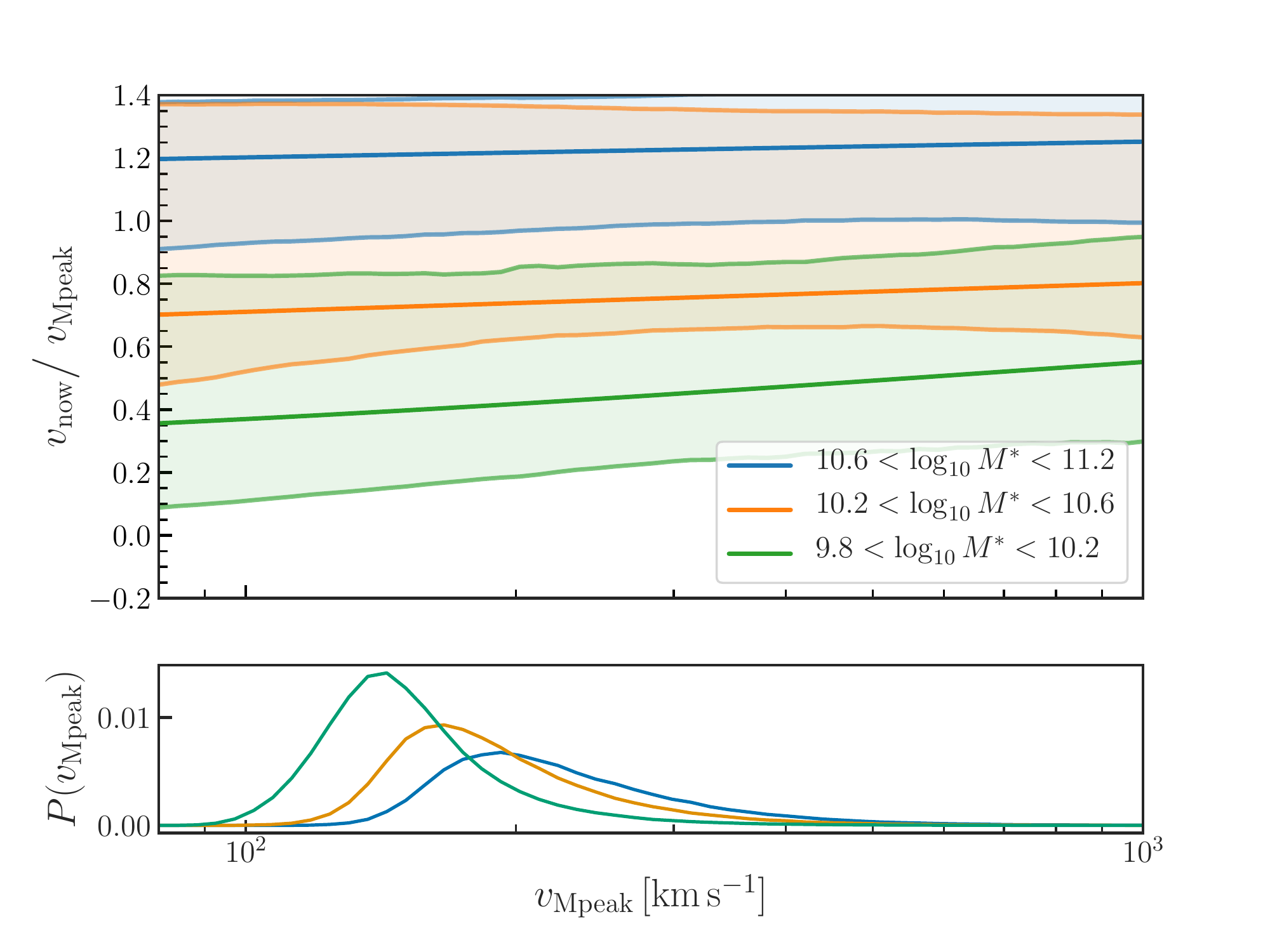} 
  \caption{({\it Top}) Constraints on $T_{\rm disr}=v_{\rm now} / \vmpeak$ as a function of \vmpeak\ for the three different stellar mass selected samples. Constraints with $T_{\rm disr} \ge 1$ imply no need for orphan galaxies. ({\it Bottom}) \vmpeak\ distribution preferred by the best fit models for each sample. The significant overlap in these distributions for the least and most massive samples, along with the different inferred $T_{\rm disr}$ for these two samples suggests that additional complexity is required in our orphan model in order to account for subhalo disruption consistently across the whole stellar mass range considered in this work.}
  \label{fig:orphan_model_constraints}
\end{figure}

Finally, we examine the \vmpeak\ dependence of our orphan model for these three different stellar mass selections in \cref{fig:orphan_model_constraints}. For the most massive bin, we again see the preference for no orphan galaxies in the fact that $T_{\rm disr}\ge 1$ for all \vmpeak, meaning that the data do not require the inclusion of any orphans for any values of \vmpeak\ populated by our model. The best fit in the second bin shows some preference for orphan galaxies, but the no-orphan scenario is not ruled out. On the other hand, for the least massive sample, we see that the mean constraint on $T_{\rm disr}$ is less than one for all values of \vmpeak, indicating that the SDSS data prefer orphan galaxies. 

We also see that there is some tension between the orphan model constraints from this least massive sample and the most massive sample, with the least massive sample requiring orphan galaxies in \vmpeak\ ranges where the most massive sample rules out orphans. The bottom panel of \cref{fig:orphan_model_constraints} shows the \vmpeak\ distributions for the three mass bins. We see that all three have significant overlap in \vmpeak\, and so the different preferences for $T_{\rm disr}$ at fixed \vmpeak\ lead to an inability to simultaneously model the least and most massive bins. This tension is also borne out by the very small Bayes factors for this model extension when considering all mass bins.

This finding suggests that there must be a secondary variable that controls artificial subhalo disruption that our model does not account for. We have tried modifying our disruption model to account for more complex dependencies, including dependence on host halo \vmpeak\ and concentration, none of which alleviate this tension. We leave more extensive investigation of these additional dependencies to future work.

\subsection{Velocity Bias}
\label{sec:velocity_bias}
We have also explored an extension to our fiducial SHAM model that allows galaxies to have different velocity distributions from the subhalo populations that they are assigned to. Such an extension is commonly referred to as velocity bias. We note that the constraints on velocity bias presented in this work are not directly comparable to constraints obtained from HOD models that place galaxies on particles or model satellite velocity distributions assuming isotropic Jeans equilibrium in an NFW profile, as the unbiased velocity distributions used in this work come from subhalo populations. Here we consider two additional parameters. The first, $\alpha_c$, allows central galaxies to have non-zero velocity with respect to the center-of-mass of the halo they are hosted by. The second parameter re-scales the velocity dispersions of satellites with respect to their host halo by a multiplicative factor, $\alpha_{s}$. The implementations of these models are described in \cref{sec:velocity_bias_model}.

\begin{figure}
\centering
  \includegraphics[width=\columnwidth]{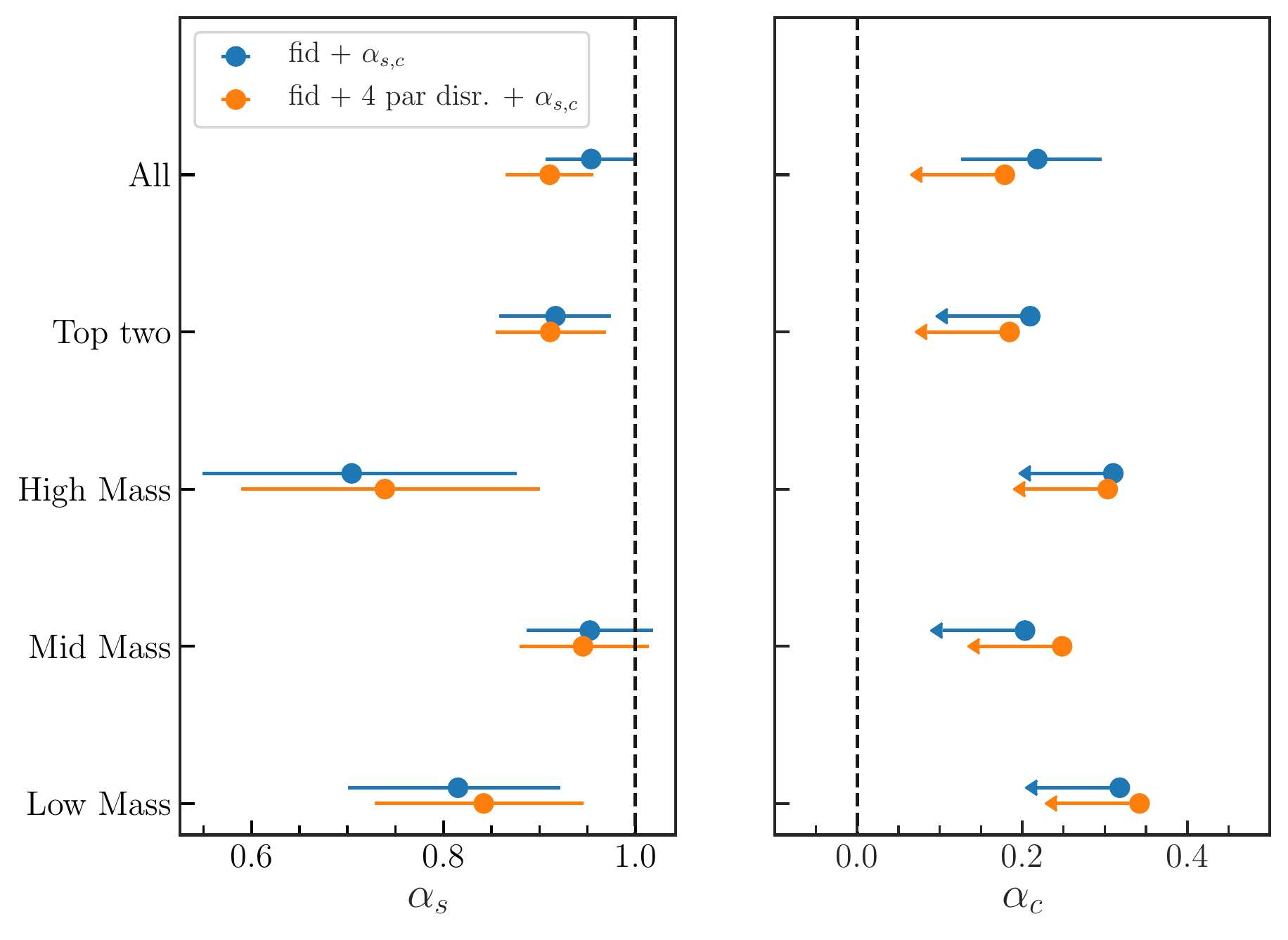}
  \caption{Same as \cref{fig:fid_model_comp}, but for velocity bias constraints. Dashed vertical lines indicate the parameter values representing no velocity bias, and error bars represent $95\%$ confidence intervals. ({\it Left}) Constraints on satellite velocity bias are inconsistent with no velocity bias at $>2\sigma$, except for combinations that include the middle mass bin. ({\it Right}) Arrows denote $2\sigma$ upper limits. Constraints on central velocity bias are all consistent with zero at $2\sigma$, except for the combined constraints from all samples when not including orphans. Including orphan galaxies does not significantly affect either satellite or central velocity bias constraints, as indicated by the consistency between the blue and orange points.}
  \label{fig:vbias_model_comp}
\end{figure}

In \cref{tab:sham_bayes_factor} we see $\mathcal{R}>1$ for model extensions that include only velocity bias other than for combinations that include the middle mass bin. \Cref{fig:vbias_model_comp} shows constraints on the velocity bias parameters for each stellar mass selection. The blue points are for model extensions including only velocity bias, and orange points show constraints when including orphan galaxies as well as velocity bias. The inclusion of orphan galaxies does not appreciably change our velocity bias constraints. It is apparent that the preference for the velocity bias model in the most and least massive galaxy samples is driven by a deviation of $\alpha_{s}$ from unity at a significance of $3.1\sigma$ and $2.6\sigma$ for the least and most massive bins respectively. When analyzing the two most massive samples simultaneously with one set of velocity bias parameters, this preference for non-unity satellite velocity bias is significantly decreased. This may indicate that the detection of velocity bias in the most massive galaxy sample is due to noise, or it may simply indicate that the $10.2 \le \log_{10}M^* < 10.6$ bin is driving the fit back towards $\alpha_{s}=1$. When all three bins are analyzed together, $\alpha_{s}$ is consistent with unity, but we caution that this result may not be robust due to the poor overall fit to the combination of all three galaxy samples. 

The central velocity bias parameter, $\alpha_c$ is consistent with 0 at $95\%$ confidence for all samples when analysed individually as shown by the blue points in in the right-hand panel of \cref{fig:vbias_model_comp}. The constraints when analyzing all samples simultaneously prefer non-zero central velocity bias at greater than two sigma but once orphan galaxies are included we find $\alpha_c$ consistent with zero for all samples.

As can be seen in \cref{fig:fid_model_comp}, including velocity bias has little effect on  parameter constraints for $\sigma_{\log_{10} M^*}$ and $\alpha$. The slight exception to this is for the $\sigma_{\log_{10} M^*}$ from the top two most massive samples, where including velocity bias changes the constraint by approximately $1\sigma$.

\section{Conditional Abundance Matching Results}
\label{sec:cam}
Here we investigate whether conditional abundance matching (CAM), is capable of modeling clustering as a function of stellar mass and specific star formation rate (SSFR). Previous sections explored SHAM extensions that may be required to model stellar mass complete galaxy selections, but such samples are rarely gathered in large spectroscopic surveys where redshift-space clustering can be measured with high precision. If CAM were capable of modeling selections in stellar mass and SSFR, then it would hold potential for modeling the clustering signals of luminous red galaxy and emission line galaxy samples observed in upcoming surveys such as DESI, 4MOST, Euclid and PFS, whose selections will implicitly depend on both stellar mass and SSFR in addition to other variables.

In the following subsections we investigate two distinct conditional abundance matching models: one that ties SSFR to the rate of change in potential well depth of subhalos as traced by their maximum circular velocities, and one that matches SSFR with distance to the nearest halo above a specified mass threshold. 

The following analyses are limited to individual stellar mass selections, as we cannot estimate a covariance matrix that is simultaneously invertible for all stellar mass and SSFR bins. As such, we cannot present CAM results using the combined constraining power of all stellar mass bins, as we did for the SHAM models. Instead, the following analyses are always based on models fit simultaneously to the quenched, star forming and combined samples for each individual stellar mass bin. As we shall see, the CAM models for individual mass bins are not entirely consistent, and so analysis of all bins in combination would have limited utility even if we could obtain a reliable covariance matrix for such an analysis.

\begin{figure*}
\centering
\includegraphics{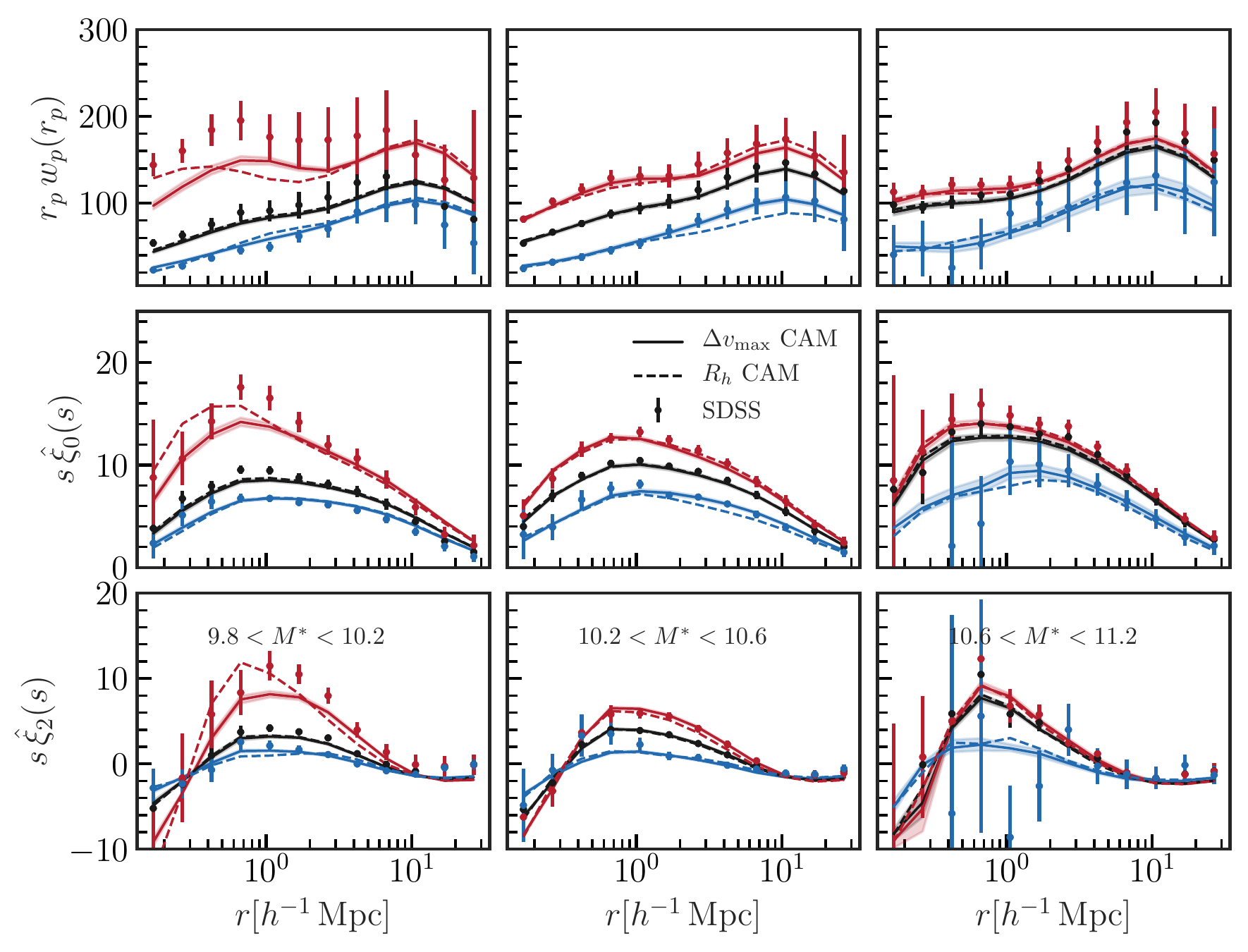}  
\caption{Comparison of the best fit \dvmax\ (solid) and \rhalo\ (dashed) CAM models to $w_p$ and \multipoles\ measured for quenched (red), star-forming (blue) and combined (black) samples for each stellar mass bin. Shaded regions around the solid lines are $1\sigma$ posterior model predictions. $1\sigma$ posterior distributions are not included around the \rhalo\ models for clarity as they are comparable to the \dvmax\ distributions in size. Overall the fits for both models are comparable, although the \dvmax\ model generally out-performs the \rhalo\ model for star-forming samples. Both models struggle to fit the quenched galaxy measurements for the two less massive bins on scales $r\le 1\,\hmpc$.}
  \label{fig:dvmax_rhalo_comp}
\end{figure*}

\begin{table*}
\caption{CAM Reduced Chi-Squared}
\begin{center}
\begin{tabular}{ c  c  c  c }
\hline
\hline
 & $9.8\le \log_{10}M^{*} < 10.2$ & $10.2\le \log_{10}M^{*} < 10.6$ &
 $10.6\le \log_{10}M^{*}$ < 11.2 \\
\hline 
Combined SHAM & 0.98 & 0.53 & 0.69\\
\hline
Combined \dvmax\ CAM & 1.13 & 0.49 & 0.82 \\
Combined \rhalo\ CAM & 1.02 & 0.49 & 0.80\\
\hline
Quenched \dvmax\ CAM& 1.76 & 0.72 & 0.61 \\
Quenched \rhalo\ CAM& 1.78 & 0.51 & 0.58 \\
\hline
Star-forming \dvmax\ CAM& 1.03 & 0.64 & 0.39 \\
Star-forming \rhalo\ CAM& 1.48 & 1.02 & 0.45 \\
\hline 
All \dvmax\ CAM& 0.62 & 0.21 & 0.18 \\
All \rhalo\ CAM& 0.52 & 0.30 & 0.18 \\
\hline 
\end{tabular}
\end{center}
\label{tab:cam_goodfit}
\end{table*}

\subsection{\dvmax\ CAM}
\label{sec:dvmax_cam}
The first CAM model that we explore is one that ties SSFR to \dvmax\ (see \cref{eq:dvmax}). This model is motivated by the success of the \texttt{UniverseMachine} in fitting the clustering and number densities of stellar mass and SSFR selected galaxies as a function of time. \texttt{UniverseMachine} also ties galaxy SSFR to \dvmax\, albeit in a model with many more free parameters in order to model star formation over a much broader range in time. Here we focus on a simplified version of the \texttt{UniverseMachine} model, where we allow for a single free parameter: the linear correlation coefficient between SSFR and \dvmax\ at fixed galaxy stellar mass. The implementation of this model is discussed in \cref{sec:cam_model}.

The solid lines in \cref{fig:dvmax_rhalo_comp} show the fits of the \dvmax\ CAM model to $w_p$, and $\hat{\xi}_{0/2}$ for quenched (red), star-forming (blue), and combined (black) samples for each stellar mass bin. Similar to \cref{fig:sham_comparison}, the fits shown in each stellar mass bin are performed independently. 

The correlations that allow the \dvmax\ CAM model to fit these signals are similar to those at play in the baseline SHAM model. For \dvmax\ CAM with $r_{\dvmax}=1$, galaxies with lower values of \dvmax\ at fixed $M^*$ are monotonically assigned galaxies with lower SSFRs. Subhalos preferentially have lower values of \dvmax\ at fixed halo mass, and so subhalos are preferentially assigned lower values of SSFR, thus boosting $f_{\rm sat}$ for more quenched samples. Additionally, \dvmax\ CAM assigns low SSFR values to the oldest host halos that are no longer accreting mass, thus boosting the central assembly bias signal in a similar way to increasing $\alpha$ in SHAM \citep{Zentner2015}. As $r_{\dvmax}$ is decreased, the correlation between SSFR and $f_{\rm sat}$, assembly bias, and, by proxy, clustering amplitude becomes flatter. Thus, $r_{\dvmax}$ governs the ratio of clustering amplitudes of the various SSFR selections at fixed $M^{*}$. 

By eye, we see that the fits to the quenched and star forming samples are reasonable, but slightly worse in general than the fits to the combined samples. This is shown more quantitatively by the reduced chi-squared values shown in \cref{tab:cam_goodfit}. The reduced chi-squared values from the $w_p + \multipoles$ row of \cref{tab:sham_results} are included for reference and listed as "Combined SHAM" indicating that the SHAM model is fit to the "combined" sample of galaxies binned in stellar mass but not SSFR. Fitting to the quenched and star-forming samples simultaneously with the combined sample degrades the fit to the combined sample for all stellar mass bins, indicating that adjustments from the best fit model from \cref{sec:base_sham} are required in order to simultaneously fit the quenched and star-forming galaxy samples. The most striking example of this is a large shift in the orphan model parameters for the least massive sample, where the \dvmax\ CAM model prefers essentially all artificially disrupted subhalos to be kept as orphans in order to fit the very large one halo term for the quenched sample in this stellar mass bin. Evidently, \dvmax\ CAM on its own cannot boost the one-halo clustering signal of this least massive bin enough on it's own. Thus, more orphan galaxies must be included in order to further boost $f_{\rm sat}$, and these orphans must be preferentially quenched. This happens naturally in \dvmax\ CAM, as orphans will be preferentially stripped and thus assigned lower SSFRs. In general, the constraints on the SHAM parameters are also broadened when fit simultaneously with $r_{\dvmax}$, again indicating a slight tension in the CAM model when fit to the data presented here. 

\begin{figure}
\centering
  \includegraphics[width=\columnwidth]{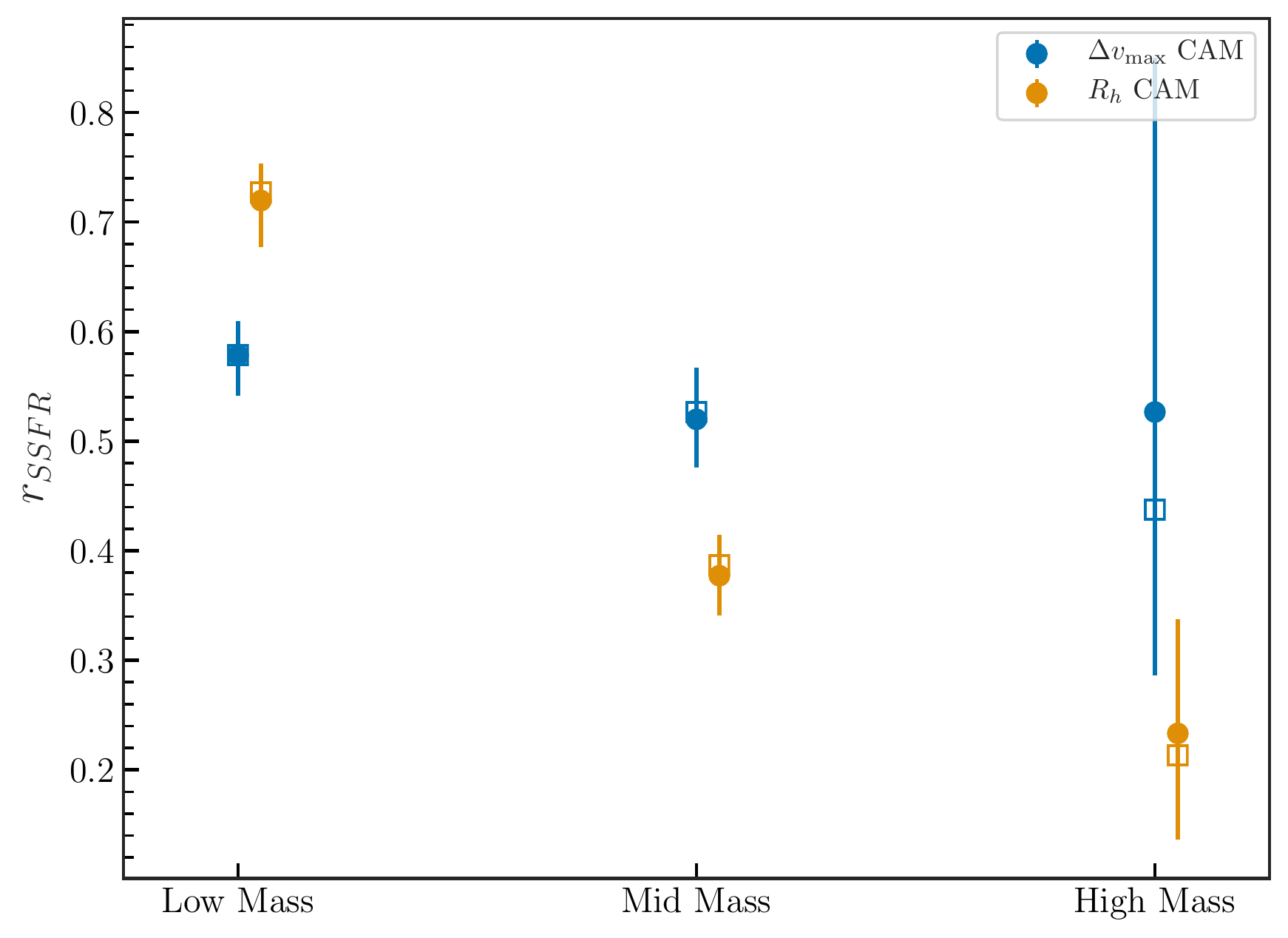}
  \caption{Posteriors of the linear correlation coefficient, $r_{SSFR}$, between SSFR and the two CAM proxies, \dvmax\ (blue), and \rhalo\ (yellow) for each stellar mass bin. Except in the least massive bin, \dvmax\ is more correlated with SSFR than \rhalo, again indicating that \dvmax\ is a better proxy for SSFR. We also see that $r_{SSFR}$ varies much more strongly with stellar mass for the \rhalo\ model than the \dvmax\ model.}
  \label{fig:rcam_constraints}
\end{figure}

\Cref{fig:rcam_constraints} shows constraints on $r_{\dvmax}$ in blue for the three stellar mass bins considered here. Notably, there is little evolution of this parameter as a function of stellar mass, with the least massive bin yielding constraints of $r_{\dvmax}=0.58\pm0.01$, while the two most massive samples prefer $r_{\dvmax}=0.52\pm0.02$ and $r_{\dvmax}=0.54\pm0.13$ respectively.

\subsection{\rhalo\ CAM}
\label{sec:rhalo_cam}
The second CAM model we consider is one that ties SSFR to \rhalo, the distance to the nearest host halo with mass above a threshold \mcut\, where \mcut\ is left as a free parameter. Here, the physics that creates the differences between high and low \rhalo\ samples is quite different than in the \dvmax\ CAM case. Subhalos that are close to massive host halos and thus have low values for \rhalo\ are assigned lower values of SSFR. This again acts to boost $f_{\rm sat}$ at fixed $M^{*}$ for quenched galaxy samples, but here there is no direct correlation between halo age and quenching, thus this model will not impart central galaxy assembly bias signals in the way that the \dvmax\ CAM model does. On the other hand, this model will introduce large correlations between quenching of nearby galaxies, also known as galactic conformity \citep{Weinmann2006}, which can have a large effect on one and two halo clustering amplitudes \citep{Hearin2015}. Like the \dvmax\ CAM model, the \rhalo\ CAM model also allows the correlation coefficient between \rhalo\ and SSFR, $r_{\rhalo}$, to be free, again controlling the ratio of clustering amplitudes for quenched and star-forming samples.

The dashed lines in \cref{fig:dvmax_rhalo_comp} show the best fit \rhalo\ model to each stellar mass bin. Overall the fits are comparable to the \dvmax\ model, although with noticeably worse performance for star-forming samples. The reduced chi-squared values for these fits are displayed in \cref{tab:cam_goodfit}, where we indeed see that the \rhalo\ model is a worse fit to the star-forming samples.

\Cref{tab:cam_bayes_factors} shows Bayes factors comparing the \dvmax\ and \rhalo\ model. The \dvmax\ model is preferred by the data for the two most massive bins, although the significance of this preference is by far the greatest in the second most massive galaxy sample. The \rhalo\ model is preferred in the least massive sample, but neither model is a particularly good fit to the data for this bin. It should be noted that these Bayes factors depend on the prior that we have chosen for \mcut, and a significantly smaller prior on this parameter would result in smaller Bayes factors and less of a preference for the \dvmax\ model.

\begin{figure}
\centering
  \includegraphics[width=\columnwidth]{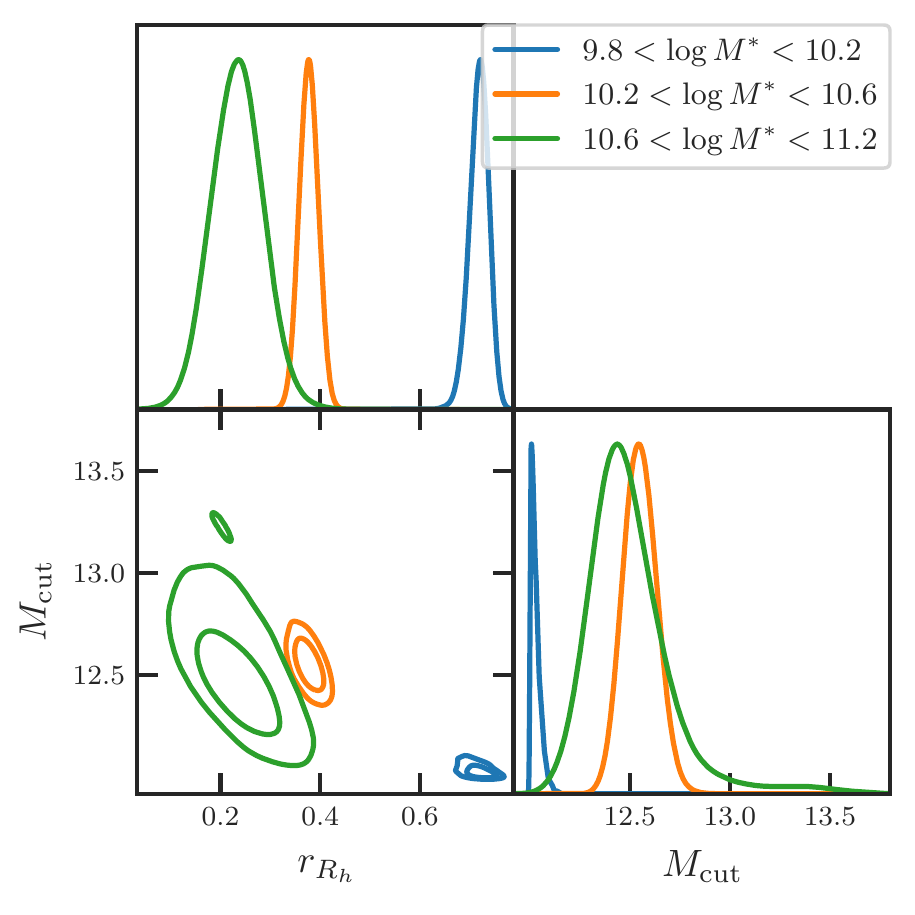}
  \caption{Posterior of the $M_{\rm cut}$ and scatter parameters used in the \rhalo\ CAM model for each stellar mass bin. $r_{\rhalo}$ and \mcut\ are strongly correlated for all samples, and both parameters show significant trends with the stellar mass of the samples. More massive samples prefer less correlation between \rhalo\ and SSFR, while also preferring larger values for \mcut.}
  \label{fig:rhalo_constraints}
\end{figure}

Joint constraints on the \rhalo\ CAM model parameters are shown in \cref{fig:rhalo_constraints}, where a strong correlation between $r_{\rhalo}$ and \mcut\ can be seen for all stellar mass bins. Unlike with the \dvmax\ model, there is a strong trend in the correlation between \rhalo\ and SSFR as a function of stellar mass. There is also an apparent trend in \mcut\ with stellar mass. More massive samples tend to be quenched by their proximity to more massive halos, as indicated by the larger preferred \mcut\ values as a function of stellar mass. The 1-dimensional posteriors on $r_{\rhalo}$ are compared with those obtained from the \dvmax\ CAM model in \cref{fig:rcam_constraints}.

\begin{table*}[htb]
\caption{CAM Bayes Factors}
\begin{center}
\begin{tabular}{ c  c  c }
\hline
\hline
$9.8\le \log_{10}M^{*} < 10.2$ & $10.2\le \log_{10}M^{*} < 10.6$ &
 $10.6\le \log_{10}M^{*}$ < 11.2 \\
\hline 
$5.8\times 10^{-6} \pm 1.7\times 10^{-6}$ & $9.9\times10^{7} \pm 3.1\times10^{7}$  & $21 \pm 5.2$\\
\hline 
\end{tabular}
\end{center}
\label{tab:cam_bayes_factors}
\end{table*}

\subsection{Model predictions for finer color selections}
Given the comparable ability of both CAM models to fit the clustering statistics considered thus far, we now look more closely at the clustering of samples selected more finely in SSFR-$M^*$ space. To do so, we divide the SDSS galaxies into four equal quartiles in SSFR, independent of $M^{*}$, yielding four bins in SSFR:
\begin{align}
    Y_{i+1} \ge \log_{10} {\rm SSFR} > Y_{i},
\end{align}
with $Y_{i} \in \{-99, -11.84, -10.98, -10.072, -6.39\}$. It should be noted that the four bins in SSFR do not contain equal numbers of galaxies for each stellar mass bin because the SSFR bin edges are determined using all galaxies with $9.8 \le \log_{10}M^{*} < 11.2$. We use the same $M^{*}$ and redshift bins as those discussed in \cref{sec:obs}. Errors on the data are again estimated via jackknife, and the simulation errors are estimated using the jackknife procedure discussed in \cref{sec:sims}.

\begin{figure*}
\centering
  \includegraphics[width=\textwidth]{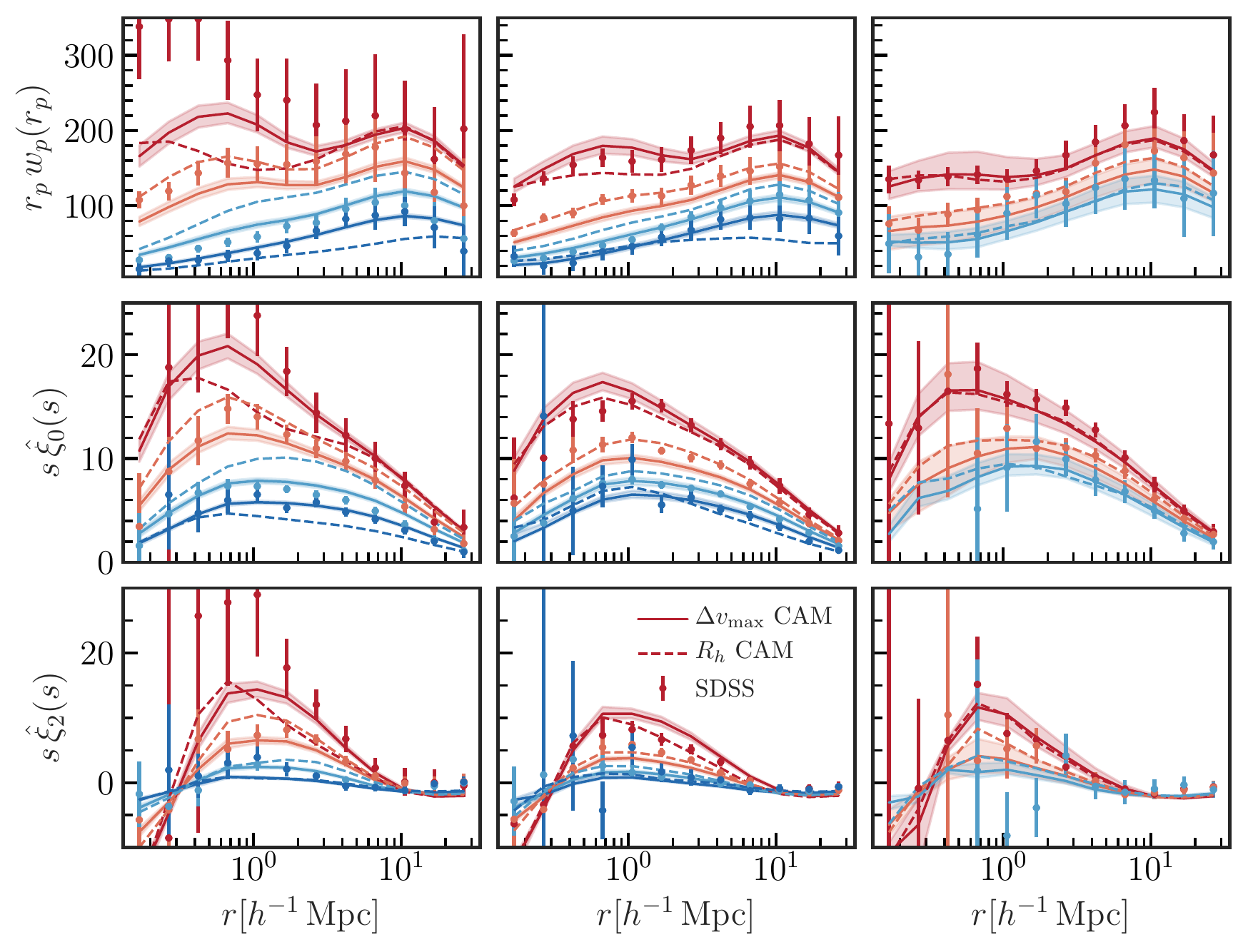}
  \caption{Predictions of the CAM models that provide the best fit to the combination of $w_p$, $\hat{\xi}_0$ and $\hat{\xi}_2$ measurements using a finer SSFR selection. Lower SSFR selections are represented by redder points and lines, while higher SSFR selections are represented by bluer points and lines. The highest SSFR sample for the most massive bin is not shown as it is extremely noisy. We see that the \dvmax\ model outperforms the \rhalo\ model for most samples, especially for the most star-forming and most quenched samples in the two less massive bins.}
  \label{fig:quartile_predictions}
\end{figure*}

Instead of fitting our CAM models to these new finer SSFR selections, we make predictions for these selections given the MAP CAM parameters presented in the previous section using the joint fit to $w_p$, $\hat{\xi}_0$ and $\hat{\xi}_2$ for each stellar mass bin. These predictions are shown in \cref{fig:quartile_predictions}, where the points represent the SDSS measurements, and the solid lines and contours are the best fit and $1\sigma$ posterior of the \dvmax\ CAM model, while dashed lines are the MAP \rhalo\ model. 

Given the small differences in the performance of these models when fit to the quenched and star-forming samples, the marked difference in the predictions for this finer binning in SSFR is surprising. For the least massive sample, the \dvmax\ model significantly out-performs the \rhalo\ model, with the \rhalo\ model predicting a very small difference in clustering amplitude for the lower SSFR samples, and a very large difference in clustering amplitude for the higher SSFR samples. These predictions average out to be roughly correct when performing the binary splits in SSFR, but these finer splits make it clear that the reasonable fit to the binary split samples is merely coincidental. For the two most massive bins, the predictions for the \dvmax\ and \rhalo\ model are more comparable, with the \dvmax\ model outperforming the \rhalo\ model for star-forming samples, while the \rhalo\ model is a better fit to the third lowest SSFR bin.

In the two less massive bins, the large difference in clustering amplitude for the two lower SSFR bins is interesting in its own right, suggesting that there is a diversity in the quenched galaxy population that is explained to a large extent by diversity in \dvmax. This can be compared to the two most star forming samples, whose clustering amplitudes are much more comparable, suggesting less diversity of environments when galaxies are on the star-forming main sequence than for quenched galaxies. This further suggests that that the data may prefer a model with separate correlation coefficients between SSFR and \dvmax\ for quenched and star-forming galaxies, with \dvmax\ being more correlated with SSFR in the quenched regime than the star-forming regime.

\subsection{Environmental Dependence of Galaxy Quenching}
\label{sec:quenching}
Because galaxy auto-correlation functions up-weight regions of high galaxy density, they are largely sensitive to the clustering of satellite galaxies, which preferentially live in these regions. For this reason, the clustering measurements presented thus far are not particularly sensitive to the processes governing central galaxy quenching, and thus not sensitive to one of the main pitfalls of the original formulation of CAM: its inclusion of very strong environmental dependence on quenching of low-mass isolated galaxies.

\citet{Tinker2017} showed that for low mass galaxies the quenched fraction, $f_{q}=\frac{n_{\rm quenched}}{n_{\rm all}}$, predicted by age matching depends very strongly on local galaxy overdensity, while in SDSS this dependence is much weaker. Here we perform our own comparison of similar measurements to the updated CAM models presented in this work to see if the reported tension persists. We adopt slightly different definitions than those presented in \citet{Tinker2017} because we use SSFR rather than $\textrm{D}_{4000}$, and because we have not run a group finder on our simulations in order to identify isolated galaxies. We define the quenched fraction as the ratio of the number of galaxies with $\log_{10} {\rm SSFR} \le -11$ to the total number of galaxies. We measure galaxy density, $\rho$, as the number of galaxies in an annulus in redshift-space defined by $0.5\,  \hinv\mpc <=r_p < 4\, \hinv\mpc$ and $\delta cz < 1000\, \textrm{km/s}$, where we only count galaxies that are less massive than the galaxy in question, and more massive massive than $0.3 M^{*}$, where $M^{*}$ is the mass of the galaxy around which we are counting satellites. 

\begin{figure*}
\label{fig:fq_v_rhogal_all}
  \includegraphics[width=\textwidth]{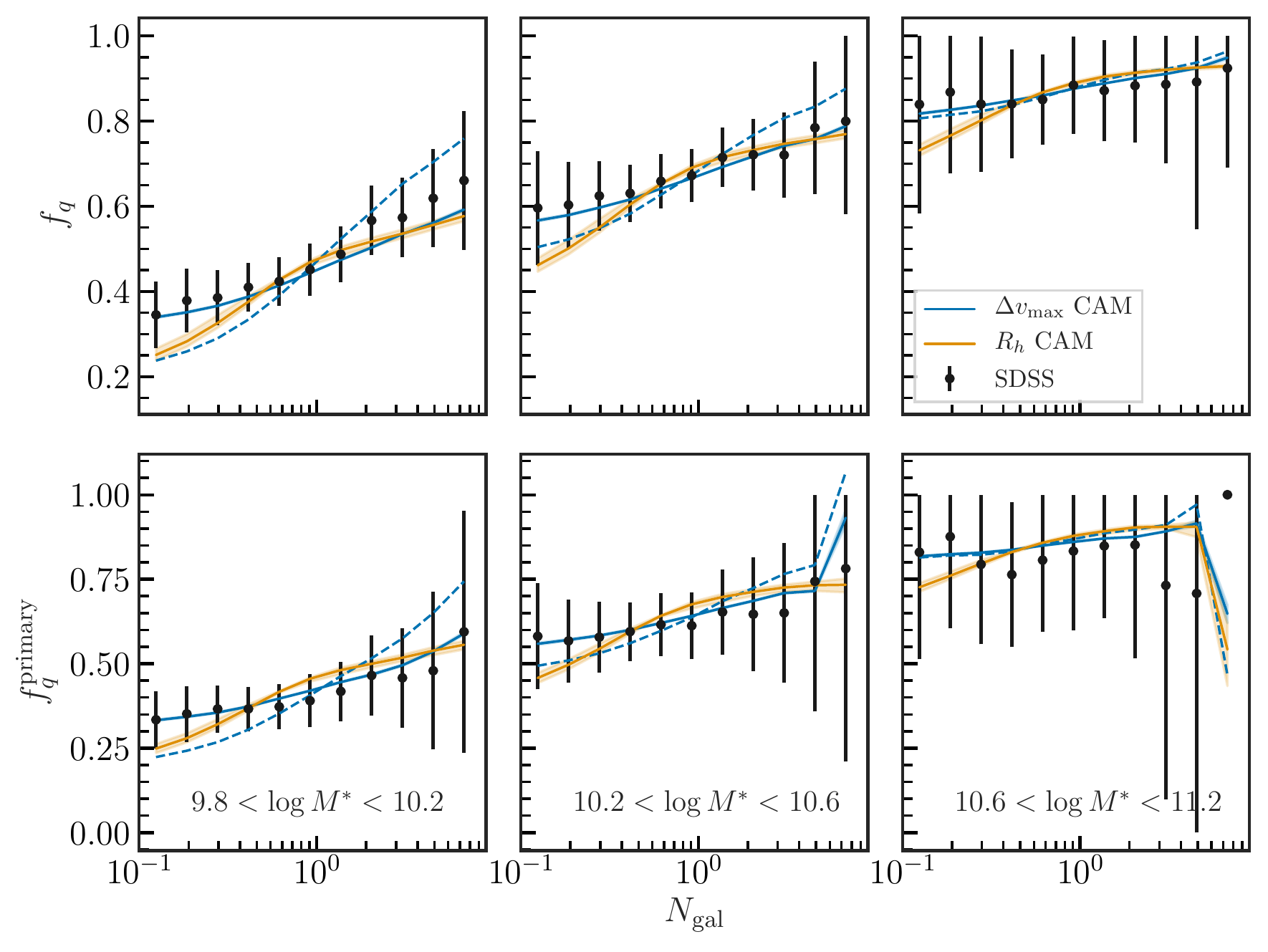}
  \caption{Predictions of quenched fraction as a function of galaxy density compared to measurements in SDSS for the \dvmax\ (blue) and \rhalo\ (orange) CAM models fit to $w_p$ and \multipoles\. The top row shows measurements of the quenched fraction of all galaxies, while the bottom row considers only primary galaxies as defined in \cref{sec:quenching}. The different columns are different stellar mass bins, as labeled in the figure. The $\Delta v_{\rm max}$ model is in good agreement with the SDSS data for all magnitudes and densities, whereas the $R_h$ model shows a significantly stronger correlation between galaxy density and $f_q$ than the data both for all galaxies, and the primary only measurements. The dashed blue line shows a \dvmax\ model with $r_{\dvmax}=1$, similar to that used in \citet{Tinker2017}, demonstrating that the inclusion of non-unity correlation between \dvmax\ and SSFR allows the data presented in this figure to be fit significantly better than the perfect correlation case.}
  \label{fig:sham_rsd}
\end{figure*}

The top row of \cref{fig:fq_v_rhogal_all} shows a comparison of $f_q$ as a function of galaxy density between our best fit CAM models from \cref{sec:cam_model} for all galaxies in each stellar mass bin. We have limited the SDSS sample's redshift range to $z<0.064$ to ensure volume completeness for all samples simultaneously, and we perform the CAM procedure using the color distributions from this sample. We see that agreement between the best fit \dvmax\ model and the data is very good for all stellar masses, while the agreement between the best fit $R_{h}$ model and the data is significantly worse, with the \rhalo\ model predicting a different shape of $f_q$ as a function of galaxy density, especially for low densities.

The bottom row of \cref{fig:fq_v_rhogal_all} shows similar measurements, but this time only considering primary galaxies, defined as those galaxies that do not have a more massive galaxy within a region of $r_p<0.5\, \hinv\mpc$ and $\Delta cz < 1000 \textrm{km}\,s^{-1}$. This is the same definition used in \citet{Behroozi2018}. We see that the trends in $f_q$ with $\rho$ become less steep than their counterparts measured for all galaxies for both the SDSS data and the CAM models. Agreement between SDSS and the $\Delta v_{\rm max}$ model is still very good for all samples, while the \rhalo\ model shows a similar shape mismatch as seen when considering all galaxies. Thus the tension reported in \citet{Tinker2017} is alleviated for the \dvmax\ model. In order to demonstrate the aspect of the updated CAM model that alleviates this tension, we also show \dvmax\ model predictions for a model with $r_{\dvmax}=1$ as dashed lines. This $r_{\dvmax}=1$ model is closer to what was used in \citet{Tinker2017}. We see that there is a much stronger trend in $f_q$ as a function of galaxy density when assuming this perfect correlation, which is very similar to that reported in \citet{Tinker2017} and in more significant tension with the SDSS measurements presented here than the best fit \dvmax model.

Recently \citet{O'Donnell2021} reported that measurements of number density profiles around isolated Milky-way mass galaxies in SDSS suggest that, if anything, star-formation rates are anti-correlated with dark matter accretion rates, seemingly contradicting the findings presented here. There are a number of differences between the measurements presented here and in \citet{O'Donnell2021}, specifically the isolation criteria used, which are more conservative in \citet{O'Donnell2021}, and the use of mean number profiles as a function of scale, rather than the counts-in-cells-like measurements that we use here. \citet{O'Donnell2021} also use accretion rates rather than \dvmax. Due to these differences it is unclear whether \citet{O'Donnell2021} contradicts our results, and further investigation confronting the statistics presented in their work with the models presented here is warranted.

Another apparent tension between CAM and SDSS data was discovered in the joint analysis of galaxy-galaxy lensing and projected clustering presented in \citet{Zu2015}. They showed that at fixed stellar mass, galaxy quenching mostly depends on halo mass, with galaxies hosted by more massive halos exhibiting stronger quenching, whereas the age-matching model discussed in their work exhibits the opposite trend, i.e. at fixed stellar mass, central galaxy quenching is inversely related to halo mass. More massive galaxies accumulate their mass at late times, and thus are younger and assigned bluer galaxies. We have investigated this tension with the best fit \dvmax\ model, and found that unlike in age matching with a perfect correlation between $z_{\rm starve}$ and SSFR, our best fit \dvmax\ model shows no signs of this inverted quenching relationship.

\section{Conclusion and Future Work}
\label{sec:conclusions}
In this work we have explored the ability of abundance matching models to fit redshift-space clustering data measured from the Sloan Digital Sky Survey Main Galaxy Sample. The parameter constraints in this work make use of polynomial chaos expansions as surrogates for abundance matching models, facilitating many aspects of our analysis. Our main findings are as follows:
\begin{enumerate}
    \item SHAM models including free parameters for scatter and the concentration dependence of the SHAM proxy can fit redshift-space clustering measurements.
    \item Orphan galaxies improve our SHAM fits, reducing the preference for \vmpeak\ abundance matching over $M_{\rm peak}$.
    \item We obtain $>2\sigma$ detections of satellite velocity bias, emphasizing the need for this degree of freedom to be included when fitting RSD statistics using SHAM.
    \item Using \dvmax\ as a proxy for SSFR in CAM models provides a good fit to the clustering data presented in this paper.
\end{enumerate}

In more detail, we fit a fiducial SHAM model with free parameters governing the scatter in stellar mass at fixed SHAM proxy, $\sigma_{\log_{10}M^*}$, and the concentration dependence of the SHAM proxy, $\alpha$, to SDSS RSD measurements in three bins of stellar mass. We find that this model is able to reproduce these measurements with high fidelity except for in the least massive galaxy sample.

\Cref{sec:disruption} and \cref{sec:velocity_bias} explore extensions to this fiducial SHAM model, allowing for orphan galaxies and velocity bias. Orphan galaxies are slightly preferred by the data for the two less massive galaxy samples, and satellite velocity bias is preferred by the least and most massive galaxy samples, as quantified by Bayes factors listed in \cref{tab:sham_bayes_factor}. Due to a lack of a consistent trend in preference or exclusion of these models, we suggest that they be included in future analyses using SHAM. Furthermore, including orphan galaxies degrades the ability to constrain $\alpha$, suggesting that previous SHAM constraints that prefer \vmpeak\ over $M_{\rm peak}$ may have been driven by artificial subhalo disruption.

While orphan galaxies allow SHAM to fit the least massive samples investigated in this work, the orphan model constraints from the three different samples are in tension. The least massive sample prefers a large number of orphans, boosting halo occupations of galaxies by a factor of $\sim 1.5$ across a broad halo mass range with respect to a no-orphan model, while the most massive sample prefers no orphans. This suggests that there are additional variables governing artificial subhalo disruption that are not included in our model, but are required in order to achieve self consistency over the full stellar mass range considered in this work.

Having investigated stellar mass complete galaxy samples, we turn to samples selected in stellar mass and SSFR, as many samples used for cosmology in ongoing and upcoming surveys will be. We present two new conditional abundance matching models, one that uses a proxy for subhalo's matter accretion rate, \dvmax, and another that uses distance to massive halos, \rhalo, as proxies for SSFR at fixed stellar mass. We have allowed for non-unity correlation between these proxies and SSFR in the form of a linear correlation coefficient. We find that both models are able to fit RSD measurements split into quenched and star-forming samples reasonably well, but the \dvmax\ model generally performs better than the \rhalo\ model. While promising, neither CAM model is able to fit the quenched and star-forming samples as well as our SHAM model is able to fit the stellar mass complete samples.

In order to perform further comparison between the CAM models, we have made predictions for additional statistics at the models' maximum \textit{a posteriori} parameter values when constrained against $w_p$, $\hat{\xi}_0$ and $\hat{\xi}_2$. First, we examined the dependence of $w_p$ and \multipoles\ in four bins of SSFR for each stellar mass bin, finding that the \dvmax\ model predictions match the SDSS measurements significantly better than the \rhalo\ model, particularly for the higher SSFR samples.

We have also examined the \textit{a posteriori} predictions of our CAM models for quenched fraction as a function of galaxy overdensity. Here, the \dvmax\ model prediction is nearly perfect and is again significantly better than the \rhalo\ model prediction, especially at low mass and low density when considering both primary and non-primary galaxies. If we force the correlation between SSFR and \dvmax\ to be perfect, then we recover the tensions between CAM and SDSS data reported in previous work, such as \citet{Tinker2017}, suggesting that allowing for non-unity correlation between SSFR and \dvmax\ is what allows us to fit this data well.

This work suggests that SHAM is a promising alternative to other simulation-based models for galaxy redshift-space clustering, particularly for higher stellar masses where only two free parameters are required to fit the SDSS data considered in this work. Modeling orphan galaxies is important for the less massive samples, but significant additional investigation is required in order to obtain an orphan model that is flexible enough to model the whole stellar mass range considered here. Including non-unity correlations in CAM alleviates many of the tensions with SDSS data that have previously been reported. CAM provides a reasonable fit to the the RSD measurements from stellar mass and SSFR selected galaxy samples, but further work is necessary in order to improve these fits to the level required for robust cosmological inference. The development of SHAM and CAM models lay the bedrock for the construction of realistic mock galaxy catalogs. With more concerted effort, SHAM and CAM may provide a robust forward model, usable for constraining the growth of structure in our universe from a broad range of galaxy clustering statistics.

\section*{Acknowledgements}
JD is supported by the Chamberlain fellowship at Lawrence Berkeley National Laboratory. The authors thank Andrew Hearin, Alexie Leauthaud, and Eduardo Rozo for useful comments, Jeremy Tinker and Zhongxu Zhai for discussions in early stages of this work, and Peter Behroozi for graciously supplying an orphan catalog for the \texttt{SMDPL} simulation. 

This work received support from the U.S. Department of Energy under contract number DE-AC02-76SF00515 at SLAC National Accelerator Laboratory. JD is supported by the Chamberlain Fellowship at Lawrence Berkeley National Laboratory. Argonne National Laboratory's work was supported by the U.S. Department of Energy, Office of Science, Office of Nuclear Physics, under contract DE-AC02-06CH11357.

The authors gratefully acknowledge the Gauss Centre for Supercomputing e.V. (www.gauss-centre.eu) and the Partnership for Advanced Supercomputing in Europe (PRACE, www.prace-ri.eu) for funding the MultiDark simulation project by providing computing time on the GCS Supercomputer SuperMUC at Leibniz Supercomputing Centre (LRZ, www.lrz.de).

The Bolshoi simulations have been performed within the Bolshoi project of the University of California High-Performance AstroComputing Center (UC-HiPACC) and were run at the NASA Ames Research Center. We are grateful to Stuart Marshall and the rest of the SLAC computing team for extensive support of this work.

\begin{appendices}
\crefalias{section}{appendix}
\section{Surrogate modeling}
\label{app:emu}

In order to facilitate our analyses, we build surrogates for the SHAM and CAM models considered in this work. In particular, we have a set of measurements, $Y=\{ y_1, ..., y_N \} \in \mathbb{R}$, at points $\mathbf{x} = \{\mathbf{x_{1}},...,\mathbf{x_{N}}\} \in \mathbb{R}^{D}$ over a domain $\mathcal{D}_{X}$, and some mapping $f: X \to Y$. The mapping $f$ is the SHAM/CAM model predictions for our clustering statistics when applied to SMDPL, and is slow to evaluate. We wish to construct a fast and accurate surrogate model $\hat{f}$ for $f$. In order to do so we construct a Polynomial Chaos Expansion for $f$ \citep{Xiu2010}. If the points in $\mathbf{X}$ follow a joint probability density function $p(\mathbf{x})$, and $f$ is a finite variance model, i.e. 
\begin{align}
    \mathbb{E}[f^2] = \int_{\mathcal{D}_X} f^2(\mathbf{x})p(\mathbf{x})d\mathbf{x} < \infty
\end{align}
and $f$ is drawn from a Hilbert space of finite variance functions, $\mathcal{H}$, endowed with the inner product
\begin{align}
    \langle g, h \rangle = \int_{\mathcal{D}_X} g(\mathbf{x})h(\mathbf{x})p(\mathbf{x})d\mathbf{x}\, ,
\end{align}
then we can construct a PCE of $f$ such that
\begin{align}
    f(\mathbf{X}) = \sum_{\mathbf{\alpha} \in \mathbb{N}^d} m_{\alpha} \Psi_{\mathbf{\alpha}}(\mathbf{X})
\end{align}
where $\mathbf{\alpha} = \{\alpha_{1},...,\alpha_{d}\}$ is a $d$-dimensional index and $\Psi_{\mathbf{\alpha}}$ is an element of the multivariate orthonormal polynomial basis of $\mathcal{H}$. 

In order to evaluate the expansion in an efficient manner it must be truncated in order to yield a finite sum, yielding an approximation to $f$: 
\begin{align}
    \hat{f}(\mathbf{X}) = \sum_{\mathbf{\alpha} \in \mathcal{A}} m_{\mathbf{\alpha}}\Psi_{\mathbf{\alpha}}(\mathbf{X})\, ,
\end{align}
where $\mathcal{A} \subset \mathbb{N}$. In this work, we choose a truncation rule such that $\mathcal{A} = \{\mathbf{\alpha} \big{|}|\mathbf{\alpha}| < p\}$, i.e. we keep all polynomials with total order less than $p$, and we optimize $p$ to compromise between accuracy and computational efficiency. For finite $p$, the PCE coefficients can be determined by optimizing a specified loss function. In our case, we take our loss function to be the mean squared error over all $\mathbf{X}$, in which case the coefficients $m_{\mathbf{\alpha}}$ are determined by solving the normal equation. 

We build independent PCEs for each scale, statistic, galaxy sample and model. In order to determine coefficients for each PCE we take $\mathbf{X}$ to be 1000 points in SHAM parameter space drawn from a Latin Hypercube. We have found that 1000 points is sufficient to achieve an accuracy consistent with the errors on the measurements in our simulations for all the models considered in this work, as discussed below. 

We populate SMDPL at each point $\mathbf{x_i}\in \mathbf{X}$ and measure $w_p(r_p)$, $\hat{\xi}_{0}(s)$ and $\hat{\xi}_{2}(s)$ as outlined in \cref{sec:sim_measurements}. This produces 1000 training points $y_i\in Y$ for each scale, statistic, galaxy sample and model. Given this data, we determine unique sets of coefficients $m_{\alpha}^{\beta}$ for each PCE, where $\beta$ now denumerates the scale, statistic, sample and model under consideration. In order to determine the generalization error of our PCEs to points outside of $\mathbf{X}$ we perform a cross-validation procedure, iteratively leaving out each point $x_{i}\in\mathbf{X}$, refitting $m_{\alpha}^{\beta}$, and calculating the residual of the PCE at the removed point, $y_{i} - \hat{f(x_{i})}$. 

\begin{figure}
\centering
      \includegraphics[width=\columnwidth]{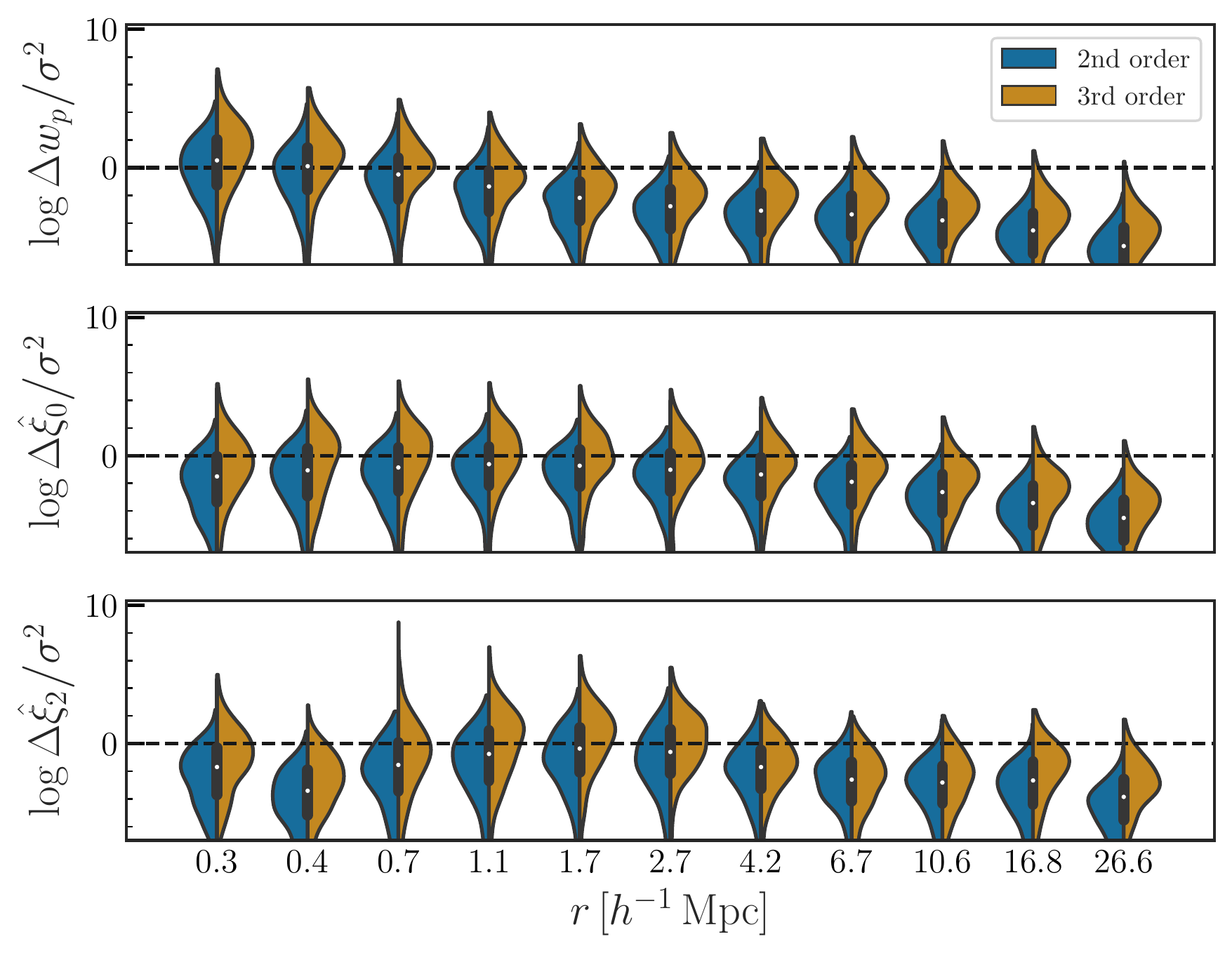}
  \caption{Test of accuracy of the emulator compared to sample variance of SMDPL. Each panel shows the distribution of emulator residual divided by estimated sample variance in SMDPL for each radial bin in $w_p$, $\hat{\xi}_{0}$ and $\hat{\xi}_{2}$ in the top, middle and bottom panels respectively. Distributions are constructed by performing leave-one-out tests where, each point in our Latin Hypercube is left out one at a time as described in \cref{sec:models}. The mean residual (white point in the violin) is smaller than sample variance ($\log_{10} \chi^2\le0$), for all but the two smallest $r_p$ bins for $w_p$ showing that our emulator is performing at the level of sample variance in our simulations or better. The blue and green distributions represent the residuals for a 2nd and 3rd order PCE respectively. We see that the 2nd order expansion is sufficient to model this data and actually outperforms the 3rd order expansion.}
  \label{fig:emulator_error}
\end{figure}

\Cref{fig:emulator_error} shows the residual distribution of the PCEs for all scales and statistics averaged over the three different stellar mass bins considered in this work. We only show the residuals for PCEs trained on the model with the greatest number of parameters, i.e. the \rhalo\ CAM model with velocity bias and subhalo disruption discussed in \cref{sec:rhalo_cam}, as we have found the residuals for this model to be the largest of all models considered in this work. Residuals are quoted as fractions of the error on each measurement, where the errors are determined via the jackknifing procedure outlined in \cref{sec:sim_measurements}. The mean error at each point, shown as the white point in each violin, is consistent with the error on the measurement for all but the two smallest radial bins in $w_p(r_p)$. We compare 2nd and 3rd order truncation schemes, and find that the second order scheme is sufficient to produce residuals consistent with the error on our training set, and so we use this scheme for all analyses in this paper. 

\section{Absolute magnitude and color selected samples}
\label{app:absmag}
In addition to the stellar mass and SSFR selected samples discussed in the rest of this work, we have also analyzed samples selected by $z=0.1$-frame $r$-band absolute magnitude and $g-r$ color. The $M_r$ selections that we use are listed in \cref{tab:sdss_absmag_samples}, and we divide red and blue samples according to \begin{align}
    \label{eq:gr_cut}
    g-r > 0.21 - 0.03\,M_r.
\end{align}

\begin{figure*}
\centering
      \includegraphics[width=\textwidth]{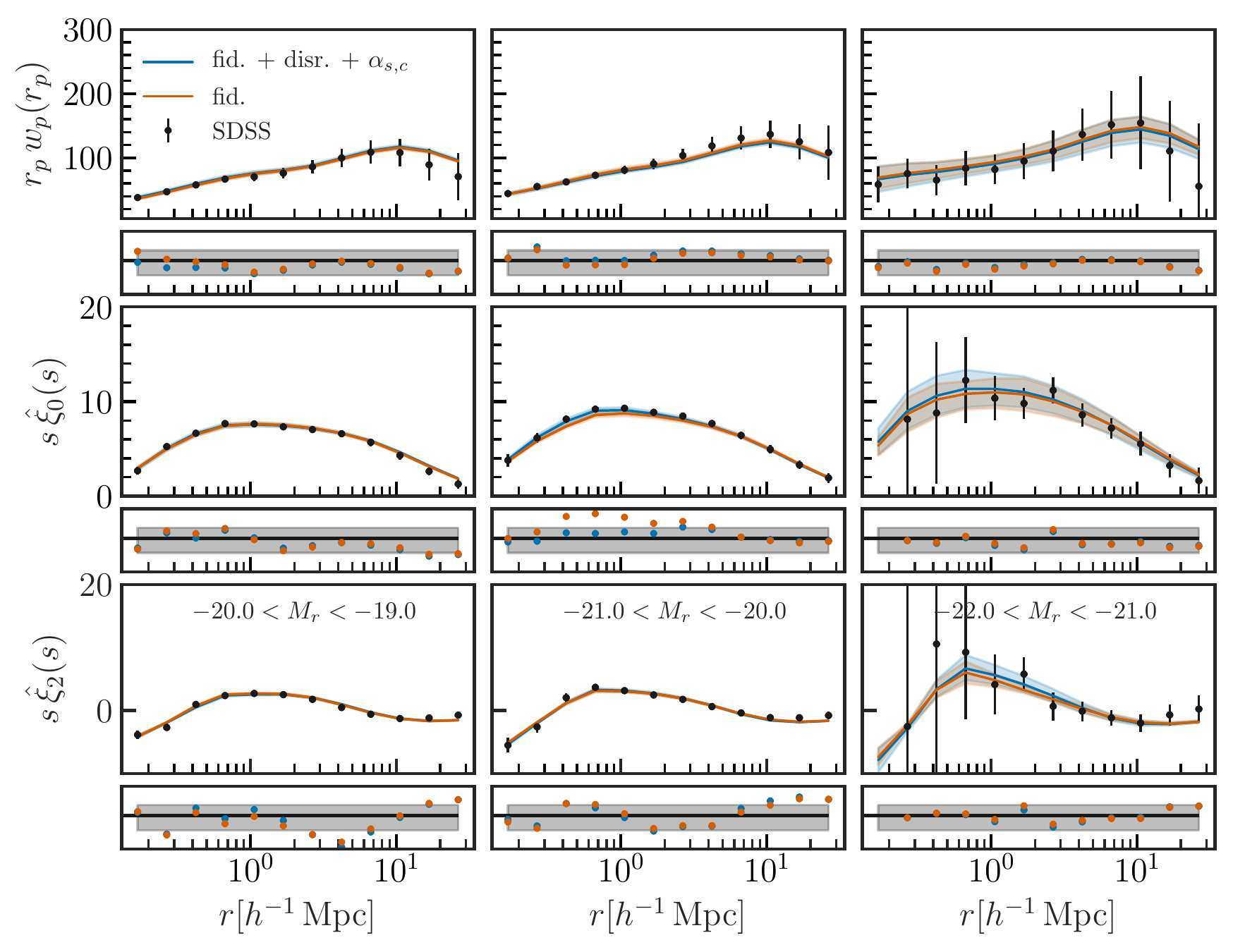} 
  \caption{Same as \cref{fig:sham_comparison}, but fitting to samples selected by $M_r$ instead of $M^{*}$. 
  Here, the models perform comparably to the stellar mass selected case although with slightly worse reduced chi-squared values for the fainter two samples. None of the model extensions are preferred by Bayes factors, but there is a significant improvement in chi-squared for the extended model in the second $M_r$ bin.}
  \label{fig:absmag_sham_comparison}
\end{figure*}

We make use of the absolute magnitude function described in \citet{Wechsler2021} for our SHAM models, but otherwise the models are implemented identically to the descriptions in \cref{sec:models}. \Cref{fig:absmag_sham_comparison} shows the fiducial and extended SHAM model fits to each absolute magnitude bin. We see similar trends in these fits to those discussed in \cref{sec:base_sham}, with the two fainter bins better fit by the extended model, while the brighter bin is fit equally well by the fiducial model and the extended model, although Bayes factors do not prefer the extended models in any of the individual $M_r$ bins. The goodness of fit for the $M_r$ selected samples is overall slightly worse than for the stellar mass selected samples but qualitatively the fits are still quite reasonable.

\begin{table}[]
    \centering
    \begin{tabular}{cccccc}
    \hline
    \hline 
        $M_r$ & $z_{\rm min}$ & $z_{\rm max}$ & $N_{\rm gal}$ & $N_{\rm red}$ & $N_{\rm blue}$\\
    \hline
        -22\textrm{ to } -21 & 0.026 & 0.106  & 21338 & 15708 & 5630\\
        -21\textrm{ to } -20 & 0.026 & 0.106  & 94630 & 57155 & 37475\\
        -20\textrm{ to } -19 & 0.026 & 0.067 & 42078 & 19641 & 22437\\
    \hline
    \end{tabular}
    \caption{SDSS sample selections using $M_r$ and $g-r$ color.}
    \label{tab:sdss_absmag_samples}
\end{table}

\begin{figure*}
\centering
\includegraphics{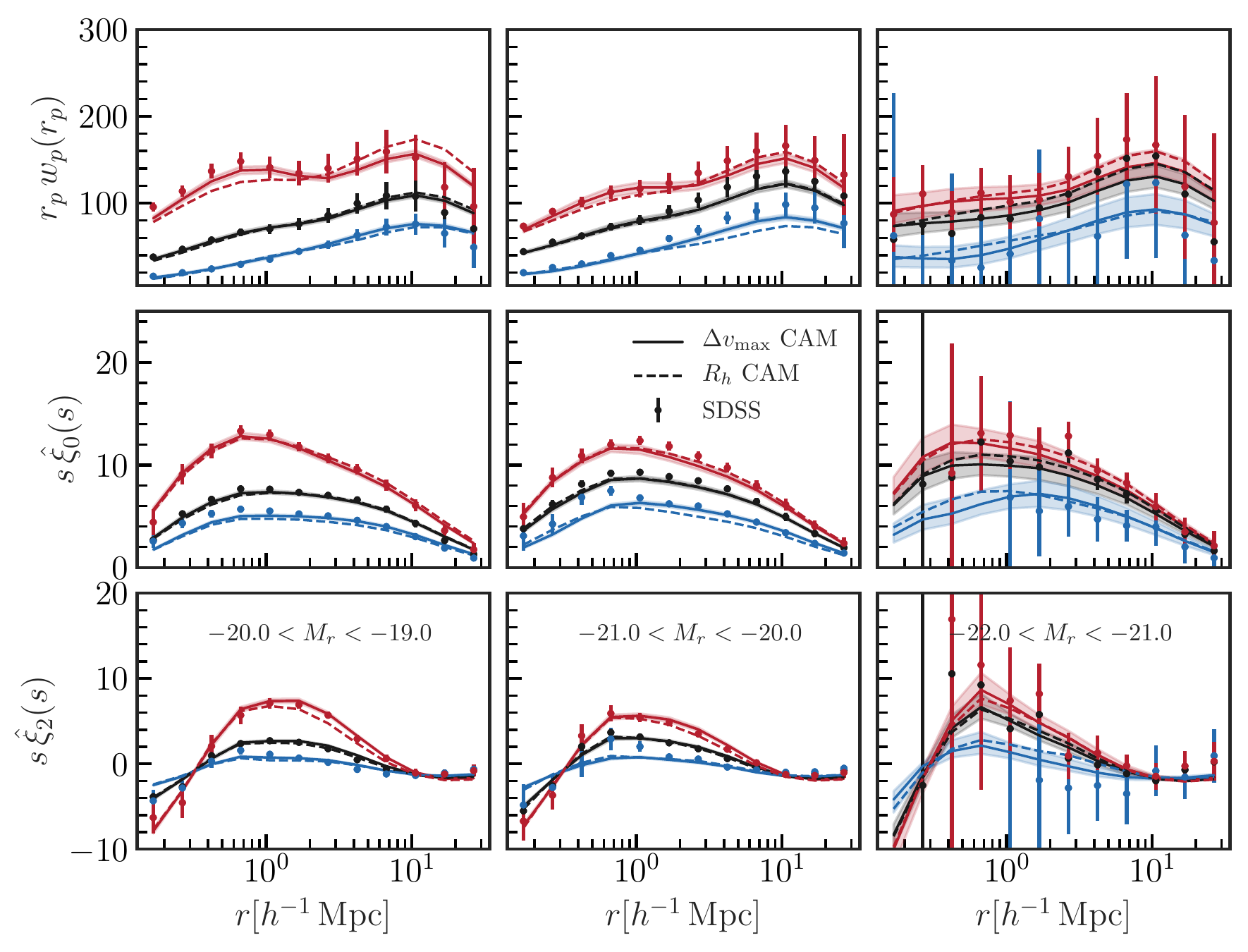}  
\caption{Same as \cref{fig:dvmax_rhalo_comp}, but for $M_r$ and $g-r$ selections, where red lines represent red galaxy selections as defined in \cref{eq:gr_cut}, and blue lines are blue galaxy selections. Overall the fits for both models are comparable, although the \dvmax\ model generally out-performs the \rhalo\ model for both blue and red galaxy samples. Unlike for the stellar mass and SSFR selections, the \dvmax\ model performs comparably across all $M_r$ bins, although with some systematic deviations for $r \le 1 \hmpc$ for blue galaxy samples.}
  \label{fig:absmag_dvmax_rhalo_comp}
\end{figure*}

Fits of the \dvmax\ and \rhalo\ CAM models to galaxy samples subdivided by $M_r$ and $g-r$ color are shown in \Cref{fig:absmag_dvmax_rhalo_comp}. Again the results are comparable to those found for the $M^{*}$ and SSFR selected samples presented in \cref{sec:cam}, with the \dvmax\ model slightly outperforming the \rhalo\ model. The main differences that are apparent are both models' ability to fit the faintest red sample for the $M_r$ selections, while the CAM models struggled to fit the analogous sample for the $M^*$ and SSFR selections. Unlike for the $M^*$ and SSFR selections, the model fits presented here cannot fit the smallest scales for the two fainter blue samples. We also find that the correlation coefficients between $g-r$ and our CAM proxies are significantly larger than in the SSFR case, suggesting that these CAM models describe the $g-r$ split data better than the SSFR selections.

\end{appendices}

\bibliographystyle{aasjournal}
\bibliography{references,aastex}

\begin{thebibliography}{}
\expandafter\ifx\csname natexlab\endcsname\relax\def\natexlab#1{#1}\fi
\providecommand{\url}[1]{\href{#1}{#1}}

\bibitem[{{Abazajian} {et~al.}(2009){Abazajian}, {Adelman-McCarthy},
  {Ag{\"u}eros}, {Allam}, {Allende Prieto}, {An}, {Anderson}, {Anderson},
  {Annis}, {Bahcall}, \& et~al.}]{Abazajian2009}
{Abazajian}, K.~N., {Adelman-McCarthy}, J.~K., {Ag{\"u}eros}, M.~A., {et~al.}
  2009, \apjs, 182, 543

\bibitem[{{Behroozi} {et~al.}(2019{\natexlab{a}}){Behroozi}, {Wechsler},
  {Hearin}, \& {Conroy}}]{Behroozi2019}
{Behroozi}, P., {Wechsler}, R.~H., {Hearin}, A.~P., \& {Conroy}, C.
  2019{\natexlab{a}}, \mnras, 488, 3143

\bibitem[{{Behroozi} {et~al.}(2019{\natexlab{b}}){Behroozi}, {Wechsler},
  {Hearin}, \& {Conroy}}]{Behroozi2018}
---. 2019{\natexlab{b}}, \mnras, 1134

\bibitem[{{Behroozi} {et~al.}(2010){Behroozi}, {Conroy}, \&
  {Wechsler}}]{Behroozi2010}
{Behroozi}, P.~S., {Conroy}, C., \& {Wechsler}, R.~H. 2010, \apj, 717, 379

\bibitem[{{Behroozi} {et~al.}(2013{\natexlab{a}}){Behroozi}, {Wechsler}, \&
  {Conroy}}]{Behroozi2013}
{Behroozi}, P.~S., {Wechsler}, R.~H., \& {Conroy}, C. 2013{\natexlab{a}}, \apj,
  770, 57

\bibitem[{{Behroozi} {et~al.}(2014){Behroozi}, {Wechsler}, {Lu}, {Hahn},
  {Busha}, {Klypin}, \& {Primack}}]{Behroozi2014}
{Behroozi}, P.~S., {Wechsler}, R.~H., {Lu}, Y., {et~al.} 2014, \apj, 787, 156

\bibitem[{{Behroozi} {et~al.}(2013{\natexlab{b}}){Behroozi}, {Wechsler}, {Wu},
  {Busha}, {Klypin}, \& {Primack}}]{Behroozi2013b}
{Behroozi}, P.~S., {Wechsler}, R.~H., {Wu}, H.-Y., {et~al.} 2013{\natexlab{b}},
  \apj, 763, 18

\bibitem[{{Berlind} \& {Weinberg}(2002)}]{BerlindWeinberg}
{Berlind}, A.~A., \& {Weinberg}, D.~H. 2002, \apj, 575, 587

\bibitem[{{Blanton} {et~al.}(2005){Blanton}, {Schlegel}, {Strauss},
  {Brinkmann}, {Finkbeiner}, {Fukugita}, {Gunn}, {Hogg}, {Ivezi{\'c}}, \&
  {Knapp}}]{Blanton2005}
{Blanton}, M.~R., {Schlegel}, D.~J., {Strauss}, M.~A., {et~al.} 2005, \aj, 129,
  2562

\bibitem[{{Bryan} \& {Norman}(1998)}]{Bryan&Norman1998}
{Bryan}, G.~L., \& {Norman}, M.~L. 1998, \apj, 495, 80

\bibitem[{{Bullock} {et~al.}(2002){Bullock}, {Wechsler}, \&
  {Somerville}}]{Bullock2003}
{Bullock}, J.~S., {Wechsler}, R.~H., \& {Somerville}, R.~S. 2002, \mnras, 329,
  246

\bibitem[{{Campbell} {et~al.}(2018){Campbell}, {van den Bosch}, {Padmanabhan},
  {Mao}, {Zentner}, {Lange}, {Jiang}, \& {Villarreal}}]{campbell2016}
{Campbell}, D., {van den Bosch}, F.~C., {Padmanabhan}, N., {et~al.} 2018,
  \mnras, 477, 359

\bibitem[{{Chaves-Montero} {et~al.}(2016){Chaves-Montero}, {Angulo}, {Schaye},
  {Schaller}, {Crain}, {Furlong}, \& {Theuns}}]{ChavesMontero2016}
{Chaves-Montero}, J., {Angulo}, R.~E., {Schaye}, J., {et~al.} 2016, \mnras,
  460, 3100

\bibitem[{{Conroy} {et~al.}(2006){Conroy}, {Wechsler}, \&
  {Kravtsov}}]{Conroy2006}
{Conroy}, C., {Wechsler}, R.~H., \& {Kravtsov}, A.~V. 2006, \apj, 647, 201

\bibitem[{{Contreras} {et~al.}(2021{\natexlab{a}}){Contreras}, {Angulo}, \&
  {Zennaro}}]{Contreras2020}
{Contreras}, S., {Angulo}, R.~E., \& {Zennaro}, M. 2021{\natexlab{a}}, \mnras,
  504, 5205

\bibitem[{{Contreras} {et~al.}(2021{\natexlab{b}}){Contreras},
  {Chaves-Montero}, {Zennaro}, \& {Angulo}}]{Contreras2021}
{Contreras}, S., {Chaves-Montero}, J., {Zennaro}, M., \& {Angulo}, R.~E.
  2021{\natexlab{b}}, arXiv e-prints, arXiv:2105.05854

\bibitem[{{Davis} \& {Peebles}(1983)}]{Davis&Peebles1983}
{Davis}, M., \& {Peebles}, P.~J.~E. 1983, \apj, 267, 465

\bibitem[{{Dawson} {et~al.}(2013){Dawson}, {Schlegel}, {Ahn}, {Anderson},
  {Aubourg}, {Bailey}, {Barkhouser}, {Bautista}, {Beifiori}, \&
  {Berlind}}]{Dawson2013}
{Dawson}, K.~S., {Schlegel}, D.~J., {Ahn}, C.~P., {et~al.} 2013, \aj, 145, 10

\bibitem[{{DeRose} {et~al.}(2019){DeRose}, {Wechsler}, {Becker}, {Busha},
  {Rykoff}, {MacCrann}, {Erickson}, {Evrard}, {Kravtsov}, {Gruen}, {Allam},
  {Avila}, {Bridle}, {Brooks}, {Buckley-Geer}, {Carnero Rosell}, {Carrasco
  Kind}, {Carretero}, {Castander}, {Cawthon}, {Crocce}, {da Costa}, {Davis},
  {De Vicente}, {Dietrich}, {Doel}, {Drlica-Wagner}, {Fosalba}, {Frieman},
  {Garcia-Bellido}, {Gutierrez}, {Hartley}, {Hollowood}, {Hoyle}, {James},
  {Krause}, {Kuehn}, {Kuropatkin}, {Lima}, {Maia}, {Menanteau}, {Miller},
  {Miquel}, {Ogando}, {Plazas Malag{\'o}n}, {Romer}, {Sanchez}, {Schindler},
  {Serrano}, {Sevilla-Noarbe}, {Smith}, {Suchyta}, {Swanson}, {Tarle}, \&
  {Vikram}}]{DeRose2018}
{DeRose}, J., {Wechsler}, R.~H., {Becker}, M.~R., {et~al.} 2019, arXiv
  e-prints, arXiv:1901.02401

\bibitem[{{DESI Collaboration} {et~al.}(2016){DESI Collaboration}, {Aghamousa},
  {Aguilar}, {Ahlen}, {Alam}, {Allen}, {Allende Prieto}, {Annis}, {Bailey},
  {Balland}, \& et~al.}]{Aghamousa2016}
{DESI Collaboration}, {Aghamousa}, A., {Aguilar}, J., {et~al.} 2016, arXiv
  e-prints, arXiv:1611.00036

\bibitem[{{Dore} {et~al.}(2019){Dore}, {Hirata}, {Wang}, {Weinberg}, {Eifler},
  {Foley}, {Heinrich}, {Krause}, {Perlmutter}, {Pisani}, {Scolnic}, {Spergel},
  {Suntzeff}, {Aldering}, {Baltay}, {Capak}, {Choi}, {Dvorkin}, {Fall}, {Fang},
  {Fruchter}, {Galbany}, {Ho}, {Hounsell}, {Izard}, {Jain}, {Koekemoer},
  {Kruk}, {Leauthaud}, {Malhotra}, {Mandelbaum}, {Massara}, {Masters},
  {Miyatake}, {Plazas}, {Rhoads}, {Rhodes}, {Rose}, {Rubin}, {Sako},
  {Samushia}, {Shirasaki}, {Simet}, {Takada}, {Troxel}, {Wu}, {Yoshida}, \&
  {Zhai}}]{Dore2019}
{Dore}, O., {Hirata}, C., {Wang}, Y., {et~al.} 2019, \baas, 51, 341

\bibitem[{{Fisher} {et~al.}(1994){Fisher}, {Davis}, {Strauss}, {Yahil}, \&
  {Huchra}}]{Fisher1994}
{Fisher}, K.~B., {Davis}, M., {Strauss}, M.~A., {Yahil}, A., \& {Huchra}, J.~P.
  1994, \mnras, 267, 927

\bibitem[{{G{\'o}rski} {et~al.}(2005){G{\'o}rski}, {Hivon}, {Banday},
  {Wandelt}, {Hansen}, {Reinecke}, \& {Bartelmann}}]{Gorski2005}
{G{\'o}rski}, K.~M., {Hivon}, E., {Banday}, A.~J., {et~al.} 2005, \apj, 622,
  759

\bibitem[{{Guo} {et~al.}(2016){Guo}, {Zheng}, {Behroozi}, {Zehavi}, {Chuang},
  {Comparat}, {Favole}, {Gottloeber}, {Klypin}, {Prada},
  {Rodr{\'\i}guez-Torres}, {Weinberg}, \& {Yepes}}]{Guo2016}
{Guo}, H., {Zheng}, Z., {Behroozi}, P.~S., {et~al.} 2016, \mnras, 459, 3040

\bibitem[{{Hartlap} {et~al.}(2007){Hartlap}, {Simon}, \&
  {Schneider}}]{Hartlap2007}
{Hartlap}, J., {Simon}, P., \& {Schneider}, P. 2007, \aap, 464, 399

\bibitem[{{Hearin} {et~al.}(2016){Hearin}, {Tollerud}, {Robitaille},
  {Droettboom}, {Zentner}, {Bray}, {Craig}, {Bradley}, {Barbary}, {Deil},
  {Tan}, {Becker}, {More}, {G{\"u}nther}, \& {Sipocz}}]{Hearin2016}
{Hearin}, A., {Tollerud}, E., {Robitaille}, T., {et~al.} 2016, {Halotools:
  Galaxy-Halo connection models}, , , ascl:1604.005

\bibitem[{{Hearin} \& {Watson}(2013)}]{Hearin2013}
{Hearin}, A.~P., \& {Watson}, D.~F. 2013, \mnras, 435, 1313

\bibitem[{{Hearin} {et~al.}(2014){Hearin}, {Watson}, {Becker}, {Reyes},
  {Berlind}, \& {Zentner}}]{Hearin2014}
{Hearin}, A.~P., {Watson}, D.~F., {Becker}, M.~R., {et~al.} 2014, \mnras, 444,
  729

\bibitem[{{Hearin} {et~al.}(2015){Hearin}, {Watson}, \& {van den
  Bosch}}]{Hearin2015}
{Hearin}, A.~P., {Watson}, D.~F., \& {van den Bosch}, F.~C. 2015, \mnras, 452,
  1958

\bibitem[{{Hearin} {et~al.}(2013){Hearin}, {Zentner}, {Berlind}, \&
  {Newman}}]{Hearin2013b}
{Hearin}, A.~P., {Zentner}, A.~R., {Berlind}, A.~A., \& {Newman}, J.~A. 2013,
  \mnras, 433, 659

\bibitem[{{Jiang} {et~al.}(2021){Jiang}, {Dekel}, {Freundlich}, {van den
  Bosch}, {Green}, {Hopkins}, {Benson}, \& {Du}}]{Jiang2020}
{Jiang}, F., {Dekel}, A., {Freundlich}, J., {et~al.} 2021, \mnras, 502, 621

\bibitem[{{Klypin} {et~al.}(2016){Klypin}, {Yepes}, {Gottl{\"o}ber}, {Prada},
  \& {He{\ss}}}]{Klypin2016}
{Klypin}, A., {Yepes}, G., {Gottl{\"o}ber}, S., {Prada}, F., \& {He{\ss}}, S.
  2016, \mnras, 457, 4340

\bibitem[{{Korytov} {et~al.}(2019){Korytov}, {Hearin}, {Kovacs}, {Larsen},
  {Rangel}, {Hollowed}, {Benson}, {Heitmann}, {Mao}, {Bahmanyar}, {Chang},
  {Campbell}, {DeRose}, {Finkel}, {Frontiere}, {Gawiser}, {Habib}, {Joachimi},
  {Lanusse}, {Li}, {Mandelbaum}, {Morrison}, {Newman}, {Pope}, {Rykoff},
  {Simet}, {To}, {Vikraman}, {Wechsler}, {White}, \& {(The LSST Dark Energy
  Science Collaboration}}]{Korytov2019}
{Korytov}, D., {Hearin}, A., {Kovacs}, E., {et~al.} 2019, \apjs, 245, 26

\bibitem[{{Kravtsov} {et~al.}(2004){Kravtsov}, {Berlind}, {Wechsler}, {Klypin},
  {Gottl{\"o}ber}, {Allgood}, \& {Primack}}]{Kravtsov2004}
{Kravtsov}, A.~V., {Berlind}, A.~A., {Wechsler}, R.~H., {et~al.} 2004, \apj,
  609, 35

\bibitem[{{Landy} \& {Szalay}(1993)}]{Landy&Szalay1993}
{Landy}, S.~D., \& {Szalay}, A.~S. 1993, \apj, 412, 64

\bibitem[{{Lange} {et~al.}(2021){Lange}, {Huntemann}, {Rahm}, {Sanner}, {Shao},
  {Lipphardt}, {Tamm}, {Weyers}, \& {Peik}}]{Lange2021}
{Lange}, R., {Huntemann}, N., {Rahm}, J.~M., {et~al.} 2021, \prl, 126, 011102

\bibitem[{{Laureijs} {et~al.}(2011){Laureijs}, {Amiaux}, {Arduini},
  {Augu{\`e}res}, {Brinchmann}, {Cole}, {Cropper}, {Dabin}, {Duvet}, {Ealet},
  \& et~al.}]{Laureijs2011}
{Laureijs}, R., {Amiaux}, J., {Arduini}, S., {et~al.} 2011, arXiv e-prints,
  arXiv:1110.3193

\bibitem[{{Leauthaud} {et~al.}(2016){Leauthaud}, {Bundy}, {Saito}, {Tinker},
  {Maraston}, {Tojeiro}, {Huang}, {Brownstein}, {Schneider}, \&
  {Thomas}}]{Leauthaud2016}
{Leauthaud}, A., {Bundy}, K., {Saito}, S., {et~al.} 2016, \mnras, 457, 4021

\bibitem[{{Lehmann} {et~al.}(2017){Lehmann}, {Mao}, {Becker}, {Skillman}, \&
  {Wechsler}}]{Lehmann2017}
{Lehmann}, B.~V., {Mao}, Y.-Y., {Becker}, M.~R., {Skillman}, S.~W., \&
  {Wechsler}, R.~H. 2017, \apj, 834, 37

\bibitem[{{Mao} {et~al.}(2018){Mao}, {Zentner}, \& {Wechsler}}]{Mao2017}
{Mao}, Y.-Y., {Zentner}, A.~R., \& {Wechsler}, R.~H. 2018, \mnras, 474, 5143

\bibitem[{{Marinacci} {et~al.}(2018){Marinacci}, {Vogelsberger}, {Pakmor},
  {Torrey}, {Springel}, {Hernquist}, {Nelson}, {Weinberger}, {Pillepich},
  {Naiman}, \& {Genel}}]{Marinacci2018}
{Marinacci}, F., {Vogelsberger}, M., {Pakmor}, R., {et~al.} 2018, \mnras, 480,
  5113

\bibitem[{{Masaki} {et~al.}(2013){Masaki}, {Lin}, \& {Yoshida}}]{Masaki2013}
{Masaki}, S., {Lin}, Y.-T., \& {Yoshida}, N. 2013, \mnras, 436, 2286

\bibitem[{{Naiman} {et~al.}(2018){Naiman}, {Pillepich}, {Springel},
  {Ramirez-Ruiz}, {Torrey}, {Vogelsberger}, {Pakmor}, {Nelson}, {Marinacci},
  {Hernquist}, {Weinberger}, \& {Genel}}]{Naiman2018}
{Naiman}, J.~P., {Pillepich}, A., {Springel}, V., {et~al.} 2018, \mnras, 477,
  1206

\bibitem[{{Nelson} {et~al.}(2018){Nelson}, {Pillepich}, {Springel},
  {Weinberger}, {Hernquist}, {Pakmor}, {Genel}, {Torrey}, {Vogelsberger},
  {Kauffmann}, {Marinacci}, \& {Naiman}}]{Nelson2018}
{Nelson}, D., {Pillepich}, A., {Springel}, V., {et~al.} 2018, \mnras, 475, 624

\bibitem[{{O'Donnell} {et~al.}(2021){O'Donnell}, {Behroozi}, \&
  {More}}]{O'Donnell2021}
{O'Donnell}, C., {Behroozi}, P., \& {More}, S. 2021, \mnras, 501, 1253

\bibitem[{{Ogiya} {et~al.}(2019){Ogiya}, {van den Bosch}, {Hahn}, {Green},
  {Miller}, \& {Burkert}}]{Ogiya2019}
{Ogiya}, G., {van den Bosch}, F.~C., {Hahn}, O., {et~al.} 2019, \mnras, 485,
  189

\bibitem[{{Peng} {et~al.}(2010){Peng}, {Lilly}, {Kova{\v{c}}}, {Bolzonella},
  {Pozzetti}, {Renzini}, {Zamorani}, {Ilbert}, {Knobel}, {Iovino}, {Maier},
  {Cucciati}, {Tasca}, {Carollo}, {Silverman}, {Kampczyk}, {de Ravel},
  {Sanders}, {Scoville}, {Contini}, {Mainieri}, {Scodeggio}, {Kneib}, {Le
  F{\`e}vre}, {Bardelli}, {Bongiorno}, {Caputi}, {Coppa}, {de la Torre},
  {Franzetti}, {Garilli}, {Lamareille}, {Le Borgne}, {Le Brun}, {Mignoli},
  {Perez Montero}, {Pello}, {Ricciardelli}, {Tanaka}, {Tresse}, {Vergani},
  {Welikala}, {Zucca}, {Oesch}, {Abbas}, {Barnes}, {Bordoloi}, {Bottini},
  {Cappi}, {Cassata}, {Cimatti}, {Fumana}, {Hasinger}, {Koekemoer},
  {Leauthaud}, {Maccagni}, {Marinoni}, {McCracken}, {Memeo}, {Meneux}, {Nair},
  {Porciani}, {Presotto}, \& {Scaramella}}]{Peng2010}
{Peng}, Y.-j., {Lilly}, S.~J., {Kova{\v{c}}}, K., {et~al.} 2010, \apj, 721, 193

\bibitem[{{Pillepich} {et~al.}(2018){Pillepich}, {Nelson}, {Hernquist},
  {Springel}, {Pakmor}, {Torrey}, {Weinberger}, {Genel}, {Naiman}, {Marinacci},
  \& {Vogelsberger}}]{Pillipich2018}
{Pillepich}, A., {Nelson}, D., {Hernquist}, L., {et~al.} 2018, \mnras, 475, 648

\bibitem[{{Planck Collaboration} {et~al.}(2016){Planck Collaboration}, {Ade},
  {Aghanim}, {Arnaud}, {Ashdown}, {Aumont}, {Baccigalupi}, {Banday},
  {Barreiro}, {Bartlett}, \& et~al.}]{Planck2015}
{Planck Collaboration}, {Ade}, P.~A.~R., {Aghanim}, N., {et~al.} 2016, \aap,
  594, A13

\bibitem[{{Reddick} {et~al.}(2013){Reddick}, {Wechsler}, {Tinker}, \&
  {Behroozi}}]{Reddick2013}
{Reddick}, R.~M., {Wechsler}, R.~H., {Tinker}, J.~L., \& {Behroozi}, P.~S.
  2013, \apj, 771, 30

\bibitem[{{Reid} {et~al.}(2014){Reid}, {Seo}, {Leauthaud}, {Tinker}, \&
  {White}}]{Reid2014}
{Reid}, B.~A., {Seo}, H.-J., {Leauthaud}, A., {Tinker}, J.~L., \& {White}, M.
  2014, \mnras, 444, 476

\bibitem[{{Saito} {et~al.}(2016){Saito}, {Leauthaud}, {Hearin}, {Bundy},
  {Zentner}, {Behroozi}, {Reid}, {Sinha}, {Coupon}, \& {Tinker}}]{Saito2015}
{Saito}, S., {Leauthaud}, A., {Hearin}, A.~P., {et~al.} 2016, \mnras, 460, 1457

\bibitem[{{Seljak}(2000)}]{Seljak2002}
{Seljak}, U. 2000, \mnras, 318, 203

\bibitem[{{Sinha} \& {Garrison}(2017)}]{Sinha&Garrison2017}
{Sinha}, M., \& {Garrison}, L. 2017, {Corrfunc: Blazing fast correlation
  functions on the CPU}, , , ascl:1703.003

\bibitem[{{Speagle}(2019)}]{Speagle2019}
{Speagle}, J.~S. 2019, arXiv e-prints, arXiv:1909.12313

\bibitem[{Springel(2005)}]{Springel2005}
Springel, V. 2005, Monthly Notices of the Royal Astronomical Society, 364,
  1105–1134.
\newblock \url{http://dx.doi.org/10.1111/j.1365-2966.2005.09655.x}

\bibitem[{{Springel} {et~al.}(2018){Springel}, {Pakmor}, {Pillepich},
  {Weinberger}, {Nelson}, {Hernquist}, {Vogelsberger}, {Genel}, {Torrey},
  {Marinacci}, \& {Naiman}}]{Springel2018}
{Springel}, V., {Pakmor}, R., {Pillepich}, A., {et~al.} 2018, \mnras, 475, 676

\bibitem[{{Tinker} {et~al.}(2017){Tinker}, {Wetzel}, {Conroy}, \&
  {Mao}}]{Tinker2017}
{Tinker}, J.~L., {Wetzel}, A.~R., {Conroy}, C., \& {Mao}, Y.-Y. 2017, \mnras,
  472, 2504

\bibitem[{{Vale} \& {Ostriker}(2004)}]{ValeOstriker2004}
{Vale}, A., \& {Ostriker}, J.~P. 2004, \mnras, 353, 189

\bibitem[{{van den Bosch} \& {Ogiya}(2018)}]{vandenbosch2018}
{van den Bosch}, F.~C., \& {Ogiya}, G. 2018, \mnras, 475, 4066

\bibitem[{{Watson} {et~al.}(2015){Watson}, {Hearin}, {Berlind}, {Becker},
  {Behroozi}, {Skibba}, {Reyes}, {Zentner}, \& {van den Bosch}}]{Watson2015}
{Watson}, D.~F., {Hearin}, A.~P., {Berlind}, A.~A., {et~al.} 2015, \mnras, 446,
  651

\bibitem[{{Wechsler} {et~al.}(2021){Wechsler}, {DeRose}, Busha, {Becker},
  Rykoff, \& Evrard}]{Wechsler2021}
{Wechsler}, R., {DeRose}, J., Busha, M., {et~al.} 2021, to be submitted to ApJ

\bibitem[{{Wechsler} {et~al.}(2006){Wechsler}, {Zentner}, {Bullock},
  {Kravtsov}, \& {Allgood}}]{Wechsler2006}
{Wechsler}, R.~H., {Zentner}, A.~R., {Bullock}, J.~S., {Kravtsov}, A.~V., \&
  {Allgood}, B. 2006, \apj, 652, 71

\bibitem[{{Weinmann} {et~al.}(2006){Weinmann}, {van den Bosch}, {Yang}, \&
  {Mo}}]{Weinmann2006}
{Weinmann}, S.~M., {van den Bosch}, F.~C., {Yang}, X., \& {Mo}, H.~J. 2006,
  \mnras, 366, 2

\bibitem[{Xiu(2010)}]{Xiu2010}
Xiu, D. 2010, Numerical methods for stochastic computations: a spectral method
  approach (Princeton university press)

\bibitem[{{Yamamoto} {et~al.}(2015){Yamamoto}, {Masaki}, \&
  {Hikage}}]{Yamamoto2015}
{Yamamoto}, M., {Masaki}, S., \& {Hikage}, C. 2015, arXiv e-prints,
  arXiv:1503.03973

\bibitem[{{Zehavi} {et~al.}(2011){Zehavi}, {Zheng}, {Weinberg}, {Blanton},
  {Bahcall}, {Berlind}, {Brinkmann}, {Frieman}, {Gunn}, {Lupton}, {Nichol},
  {Percival}, {Schneider}, {Skibba}, {Strauss}, {Tegmark}, \&
  {York}}]{Zehavi2011}
{Zehavi}, I., {Zheng}, Z., {Weinberg}, D.~H., {et~al.} 2011, \apj, 736, 59

\bibitem[{{Zentner} {et~al.}(2014){Zentner}, {Hearin}, \& {van den
  Bosch}}]{Zentner2015}
{Zentner}, A.~R., {Hearin}, A.~P., \& {van den Bosch}, F.~C. 2014, \mnras, 443,
  3044

\bibitem[{{Zhai} {et~al.}(2019){Zhai}, {Tinker}, {Becker}, {DeRose}, {Mao},
  {McClintock}, {McLaughlin}, {Rozo}, \& {Wechsler}}]{Zhai2018}
{Zhai}, Z., {Tinker}, J.~L., {Becker}, M.~R., {et~al.} 2019, \apj, 874, 95

\bibitem[{{Zu} \& {Mandelbaum}(2016)}]{Zu2015}
{Zu}, Y., \& {Mandelbaum}, R. 2016, \mnras, 457, 4360

\end{thebibliography}

\end{document}